
\input apj2col
\hsize=7.0in
\hoffset=-0.35in
\def\etal{et al.~}

\journal{\hskip-3pt, 522:000-000, 1999 September 10}
\vskip0pt
\title{COSMIC HISTORIES OF STARS, GAS, HEAVY ELEMENTS, AND DUST IN GALAXIES}
\vskip6pt
\author{Yichuan C. Pei, S. Michael Fall, and Michael G. Hauser}
\address{Space Telescope Science Institute, 3700 San Martin
                 Drive, Baltimore, Maryland 21218}
\address{{\it Received 1998 November 5; accepted 1999 April 21}}

\vskip6pt
\abstract
We investigate a set of coupled equations that relates the stellar,
gaseous, chemical, and radiation contents of the universe averaged
over the whole population of galaxies.  Using as input the available
data from quasar absorption-line surveys, optical imaging and redshift
surveys, and the {\it COBE} DIRBE and FIRAS extragalactic infrared
background measurements, we obtain solutions for the cosmic histories
of stars, interstellar gas, heavy elements, dust, and radiation from
stars and dust in galaxies.  Our solutions reproduce remarkably well a
wide variety of observations that were not used as input.  These
include the integrated background light from galaxy counts from
near-ultraviolet to near-infrared wavelengths, the rest-frame optical
and near-infrared emissivities at various redshifts from surveys of
galaxies, the mid-infrared and far-infrared emissivities of the local
universe from the IRAS survey, the mean abundance of heavy elements at
various epochs from surveys of damped Lyman-alpha systems, and the
global star formation rates at several redshifts from H$\alpha$,
mid-infrared, and submillimeter observations.  The chemical enrichment
history of the intergalactic medium implied by our models is also
consistent with the observed mean metal content of the Lyman-alpha
forest at high redshifts.  We infer that the dust associated with star
forming regions is highly inhomogeneous and absorbs a significant
fraction of the starlight, with only $41-46$\% of the total in the
extragalactic optical background and the remaining $59-54$\%
reprocessed by dust into the infrared background.  The solutions
presented here provide an intriguing picture of the cosmic mean
history of galaxies over much of the Hubble time.  In particular, the
process of galaxy formation appears to have undergone an early period
of substantial inflow to assemble interstellar gas at $z\gtrsim3$, a
subsequent period of intense star formation and chemical enrichment at
$1\lesssim z\lesssim3$, and a recent period of decline in the gas
content, star formation rate, optical stellar emissivity, and infrared
dust emissivity at $z\lesssim1$.
\subject
cosmology: diffuse radiation --- galaxies: evolution ---
          quasars: absorption lines --- infrared: galaxies

\maintext
\section{1. INTRODUCTION}

In the past few years, galaxies have been observed to redshifts
approaching five, when the universe was less than 10\% of its present
age.  A goal of several recent studies has been to determine the
average evolution of the stellar and interstellar contents of the
entire population of galaxies.  Individual galaxies, spanning a wide
range of sizes, masses, and morphological types, certainly had
different, possibly complicated histories of star formation, gas
consumption, and chemical enrichment.  Studying the average properties
of galaxies has the virtue of simplicity and is relevant to several
important problems, such as the extragalactic background radiation and
the chemical enrichment of the intergalactic medium.  Quasar
absorption-line and optical imaging and redshift surveys of galaxies,
when combined with cosmic infrared background measurements, promise to
reveal the unbiased global history of star formation in galaxies.
This is the main objective of this paper.

Quasar absorption lines have been used for more than a decade to probe
the interstellar content of galaxies at high redshifts.  The damped
Ly$\alpha$ systems, a subset of the absorption-line systems with
highest HI column densities, are of particular interest because they
dominate the mass density of neutral gas in the universe and probably
represent the progenitors of present-day galaxies (Wolfe \etal 1986),
although the exact nature of these objects (i.e., their morphologies,
sizes, and space density) is a topic of current speculation and
investigation.  Surveys of damped Ly$\alpha$ systems along many lines
of sight provide information on the mean comoving density of neutral
hydrogen and the mean abundances of heavy elements, dust, and
molecules at various redshifts.  The picture that emerges, while still
incomplete, already reveals several important results.  First, the
damped Ly$\alpha$ systems at $z\approx3$ contain enough cool, neutral
gas to form all of the stars visible today (Wolfe \etal 1995).
Second, the observed decrease of the gas content from $z\approx3$ to
$z=0$ plausibly reflects the conversion of gas into stars (Lanzetta,
Wolfe, \& Turnshek 1995).  Third, the damped Ly$\alpha$ systems at
$z\approx3$ are still in their infancy of chemical enrichment, with
low metallicity (Pettini \etal 1994), low dust-to-gas ratio (Pei,
Fall, \& Bechtold 1991) and few molecules (Levshakov \etal 1992).
These observations, when tied together by models of cosmic chemical
evolution, predict high rates of star formation at $1\lesssim
z\lesssim2$ and a rapid decline toward $z=0$ (Pei \& Fall 1995).

Recently, optical imaging and redshift surveys have given us a direct
view of the stellar content of the distant universe (Lilly \etal 1995;
Steidel \etal 1996; Ellis \etal 1996; Cowie \etal 1996; Williams \etal
1996).  An important quantity that can be measured from these surveys
is the mean rest-frame ultraviolet emissivity as a function of
redshift, which potentially traces the global history of star
formation in the universe.  The optical data suggest that the average
rate of star formation increases with time at $z\gtrsim2$ (Madau \etal
1996), peaks near $z\approx1-2$ (Connolly \etal 1997), and then
declines sharply to $z=0$ (Lilly \etal 1996), although the apparent
increase from $z\approx 4$ to $z\approx3$ is quite uncertain (Steidel
\etal 1999).  This picture is broadly consistent with the history of
star formation inferred from models of cosmic chemical evolution with
input from surveys of damped Ly$\alpha$ systems (Fall 1998).  Thus,
cosmic chemical evolution provides a key link between the stellar,
gaseous, and chemical constituents of galaxies.

All-sky surveys from the DIRBE and FIRAS instruments on the {\it COBE}
satellite have provided us with another view of the integrated
emission history of the universe from far-infrared to millimeter
wavelengths (Puget \etal 1996; Schlegel, Finkbeiner, \& Davis 1998;
Hauser \etal 1998; Fixsen \etal 1998).  The recent DIRBE and FIRAS
detections of the cosmic infrared background, when compared with the
Hubble Deep Field survey, indicate that a significant fraction of the
radiation released by stars has not been seen in the ultraviolet and
optical region of the extragalactic background, but rather at
far-infrared wavelengths (Hauser \etal 1998).  The large amount of
far-infrared radiation seen by the DIRBE and FIRAS appears to imply
that the global star formation rate currently deduced from optical
surveys is severely biased (Dwek \etal 1998; Blain \etal 1999;
Calzetti \& Heckman 1999).  This bias is not entirely surprising
because galaxies undergoing bursts of star formation are known not
only to be sources of intense ultraviolet radiation but also of
far-infrared radiation.  The cosmic infrared background records the
cumulative infrared emission from galaxies at all redshifts and
therefore provides an important constraint on the global history of
star formation.

In this paper, we combine the available data from quasar
absorption-line surveys, optical imaging and redshift surveys, and
extragalactic background measurements to derive the cosmic histories
of star formation, gas consumption, and chemical enrichment in
galaxies.  Our approach is an extension of the studies of Pei \& Fall
(1995) and Fall, Charlot, \& Pei (1996).  These are based on a set of
equations that link the comoving densities of stars, interstellar gas,
heavy elements, and dust, averaged over the whole population of
galaxies.  The data from quasar absorption-line surveys and optical
imaging and redshift surveys are taken as inputs, and the effects of
dust are self-consistently coupled to the production of heavy
elements.  A key constraint in the present paper is imposed by the
recent DIRBE and FIRAS measurements of the cosmic infrared background.
With a self-consistent treatment of the absorption and reradiation of
starlight by dust, the problem is then to solve the set of coupled
chemical equations so that the reradiated starlight by dust matches
the background measurements.  Section~2 presents the basic method,
states the key assumptions, and identifies the necessary observational
inputs.  In \S~3, we derive the global histories of stars,
interstellar gas, heavy elements, dust, radiation from stars and dust,
and baryon flow between the interstellar and intergalactic media.  In
\S~4, we discuss the major uncertainties in the models and some of the
implications of our results.  Our conclusions are summarized in \S~5.
Throughout the paper, we adopt a cosmology with $\Lambda=0$,
$q_0=1/2$, and $H_0=50~{\rm km~s}^{-1}~{\rm Mpc}^{-1}$.

\section{2. MODELS}

We are interested in the global evolution of stars, interstellar gas,
heavy elements, dust, and radiation from stars and dust, all averaged
in comoving volumes large enough to contain many galaxies.  The basic
formalism that can be used to describe this evolution has been
developed by Lanzetta \etal (1995), Pei \& Fall (1995), and Fall \etal
(1996).  Our previous work showed that the emission history of
galaxies could be inferred from their absorption history with the aid
of stellar population synthesis models, and that self-consistent
solutions of the equations of cosmic chemical evolution predicted high
rates of star formation at $1\lesssim z\lesssim2$ and declining rates
at $0\lesssim z\lesssim 1$.  Here, we adopt nearly the same formalism
in describing cosmic chemical evolution but without any assumption
about the baryon flow rate between the interstellar and intergalactic
media, and develop some new formulae for the absorption and
reradiation of starlight by dust.  With observations now available on
the absorption and emission histories of galaxies, we can use both of
these as inputs, while the reradiation of starlight absorbed by dust
is constrained by the measured cosmic infrared background radiation.
In this way, we obtain the unbiased absorption and emission histories
of galaxies as well as the evolution of baryons within galaxies.

\subsection{2.1. Cosmic Chemical Evolution}

The equations of cosmic chemical evolution govern the global rates of
star formation, gas consumption, and chemical enrichment.  For a
population of galaxies, these rates can be expressed most simply in
terms of the mean comoving densities of stars, interstellar gas, and
elements heavier than He in the interstellar medium, $\Omega_s$,
$\Omega_g$, and $\Omega_m$, all measured in units of the present
critical density ($\rho_c=3H^2_0/8\pi G$), and the mean abundance of
heavy elements in the interstellar media of galaxies $Z\equiv\Omega_m/
\Omega_g$.  As defined here, $\Omega_s$, $\Omega_g$, $\Omega_m$, and
$Z$ all pertain to galaxies themselves, not to the intergalactic
medium.  On the timescale of interest here (the Hubble time), any
delayed recycling of stellar material to the interstellar medium can
be neglected.  We can therefore write, in the approximation of
instantaneous recycling (and ${Z\ll1}$),
\vskip-10pt
$$
\dot\Omega_g+\dot\Omega_s=\dot\Omega_f, \eqno(1)
$$
\vskip-20pt
$$
\Omega_g\dot Z-y\dot\Omega_s=(Z_f-Z)\dot\Omega_f, \eqno(2)
$$
\vskip-6pt\noindent
where $y$ is an ``effective yield'' of heavy elements and the dots
denote differentiation with respect to proper time.  The source terms
on the right-hand sides of equations (1) and (2), with subscript $f$,
represent the inflow or outflow of gas with mean metallicity $Z_f$ at
a net comoving rate $\dot\Omega_f$ to or from galaxies.  These
equations can be derived by summing the corresponding equations for
the chemical evolution of all the individual galaxies in a large
comoving volume (see Tinsley 1980).

The definition of the effective yield requires some care.  It could be
defined as the mean nucleosynthetic yield of different galaxies
weighted by their star formation rates.  In this case, however,
equation (2) would only be valid for a population of chemically
homogeneous galaxies (i.e., all galaxies at a given redshift have the
same metallicity).  The reason for this is that the average over
galaxies leading to equation (2) contains a residual term due to the
metallicity spread of different galaxies, which can simply be absorbed
into the definition of $y$.  In this case, equation (2) is also valid
for a population of chemically inhomogeneous galaxies.  This
definition of the effective yield is entirely appropriate in our
models, because we regard $y$ as an output parameter fixed by the mean
metallicity at the present epoch (see below).

We now consider a simple case of the inflow or outflow of gas between
galaxies and their environment.  Inflow ($\dot \Omega_f \ge0$) allows
for the addition of unprocessed material ($Z_f=0$) into galaxies from
the intergalactic medium, while outflow ($\dot\Omega_f<0$) allows for
the removal of processed material ($Z_f=Z$) from galaxies by supernova
explosions, galactic winds, collisions, and so forth.  Thus, we write
\vskip-6pt
$$
Z_f=Z[1-\theta(\dot\Omega_f)], \eqno(3)
$$
\vskip-6pt\noindent
where theta is the unit step function, defined as ${\theta(x)=1}$ for
$x\ge0$ and $\theta(x)=0$ for $x<0$.  If a significant outflow occurs,
it can enrich the intergalactic medium with heavy elments.  However,
since the gas reservoir in the intergalactic medium is sufficiently
large, its overall metallicity is likely to remain lower than that of
the interstellar medium within galaxies.  Therefore, the assumption of
only metal-free inflow should remain roughly valid.  In principle,
there is a possibility that both inflow and outflow could occur at the
same time and with the same rate so that the net flow is zero, in
which case equation (3) would be invalid.  In practice, as long as the
net flow is dominated by either inflow or outflow, equation (3) should
be a good approximation.

Equations (1) and (2) can be solved by specifying the initial
conditions at $t=0$ or the boundary conditions at the present epoch,
$z=0$.  We assume that, at some early time, before galaxies form,
there are no stars, interstellar gas, or heavy elements, i.e.,
$\Omega_s=\Omega_g =\Omega_m=0$ at ${t=0}$.  With these initial
conditions and equation (3), the solutions to equations (1) and (2)
are
\vskip-6pt
$$
\Omega_f=\Omega_g+\Omega_s, \eqno(4)
$$
\vskip-20pt
$$
Z=\int_0^t dt'y{\dot\Omega_s(t')\over\Omega_g(t')}\exp\biggl[
\int_{t}^{t'}dt^{\prime\prime}{\theta[\dot\Omega_f(t^{\prime\prime})]
\dot\Omega_f(t^{\prime\prime})\over\Omega_g(t^{\prime\prime})}\biggl].
\eqno(5)
$$
\vskip2pt\noindent
The formation and early evolution of galaxies naturally requires an
initial inflow, $\dot\Omega_f>0$.  Any time variation in $y$ is likely
to be weaker than that in $\dot\Omega_s$, which varies by more than an
order of magnitude over $0\lesssim z\lesssim5$.  As an approximation,
we assume that the effective yield $y$ is constant in time.  Since $Z$
and $y$ always appear in the combination $Z/y$, we can fix $y$ by
imposing a boundary condition on the mean metallicity $Z$ at $z=0$.
The quantity $Z(0)$, which refers to an average over all galaxies at
the present epoch, is not known precisely.  We simply assume that the
metallicity in the solar neighborhood is typical and adopt $Z(0)=
Z_\odot$.  The values of $Z(0)$ and all other input parameters in our
models are listed in Table~1.

Our choice of the initial conditions for $\Omega_s$ and $\Omega_g$ has
a special meaning in the models of cosmic chemical evolution: the
quantity $\Omega_f$ at any redshift is always the sum of the comoving
densities of stars and interstellar gas, and hence represents the
comoving density of baryons within galaxies.  The evolution of the
baryon content in galaxies is regulated by the star formation rate and
the gas consumption and inflow or outflow rates.  If the star
formation rates were relatively small at some early epochs, i.e.,
$\dot\Omega_s\ll\dot\Omega_g$, the evolution of galaxies would be
dominated by the build up of interstellar gas, i.e., $\dot\Omega_f
\approx \dot\Omega_g$.  On the other hand, if the star formation rates
were relatively high, this would dominate the evolution.  In fact, the
process of galaxy formation might proceed through two phases: (1) the
accretion of baryons during an early period of gas assembly, and (2)
the production of heavy elements and consumption of gas during a
subsequent period of star formation.

The total abundance of heavy elements $Z$ in equation (5) includes
metals in both the gas and solid phases of the interstellar medium.
The production of dust is therefore included naturally in our
approach.  We assume for simplicity that a fixed fraction of the heavy
elements in the interstellar medium is locked up in dust grains.  In
this case, the comoving density of dust $\Omega_d$ is related to the
comoving density of heavy elements $\Omega_m$ by
\vskip-6pt
$$
\Omega_d=d_m\Omega_m=d_m Z\Omega_g, \eqno(6)
$$
\vskip-6pt\noindent
with a constant dust-to-metals mass ratio $d_m$.  In addition to some
theoretical expectations for a constant $d_m$ (Dwek 1998), there is
also empirical evidence for this in nearby galaxies and the damped
Ly$\alpha$ systems at high redshifts.  For the Milky Way and the Large
and Small Magellanic Clouds, the dust-to-metals ratio is 0.51 (GAL),
0.36 (LMC), and 0.46 (SMC) (Pei 1992; Luck \& Lambert 1992), roughly
constant even though the dust-to-gas ratios and metallicities in these
galaxies vary by nearly an order of magnitude.  Using the measurements
of the gas-phase depletion of Cr relative to Zn in a sample of damped
Ly$\alpha$ systems by Pettini \etal (1997a), we derive mean values
$d_m=0.45$ at $0.7<z\le1.0$, 0.37 at $1.0<z\le 2.0$, and 0.44 at
$2.0<z\le2.8$, with the assumptions that the intrinsic Cr/Zn is solar,
that Cr is depleted onto grains, while Zn is not, and that the dust
has Galatic chemical composition and SMC-type extinction curve.  These
results are consistent with estimates based on the observed gas-phase
depletions of other elements in damped Ly$\alpha$ systems (Kulkarni,
Fall, \& Truran 1997; Vladilo 1998).  Figure 1 shows our estimates of
$d_m$ as a function of redshift, where the data point at $z=0$ is the
mean for the Milky Way, LMC, and SMC.  Evidently, the mean
dust-to-metals ratio is roughly constant at $d_m=0.45$ over a
significant range of redshifts, $0\le z\lesssim3$.

\vskip20pt
\psfig{file=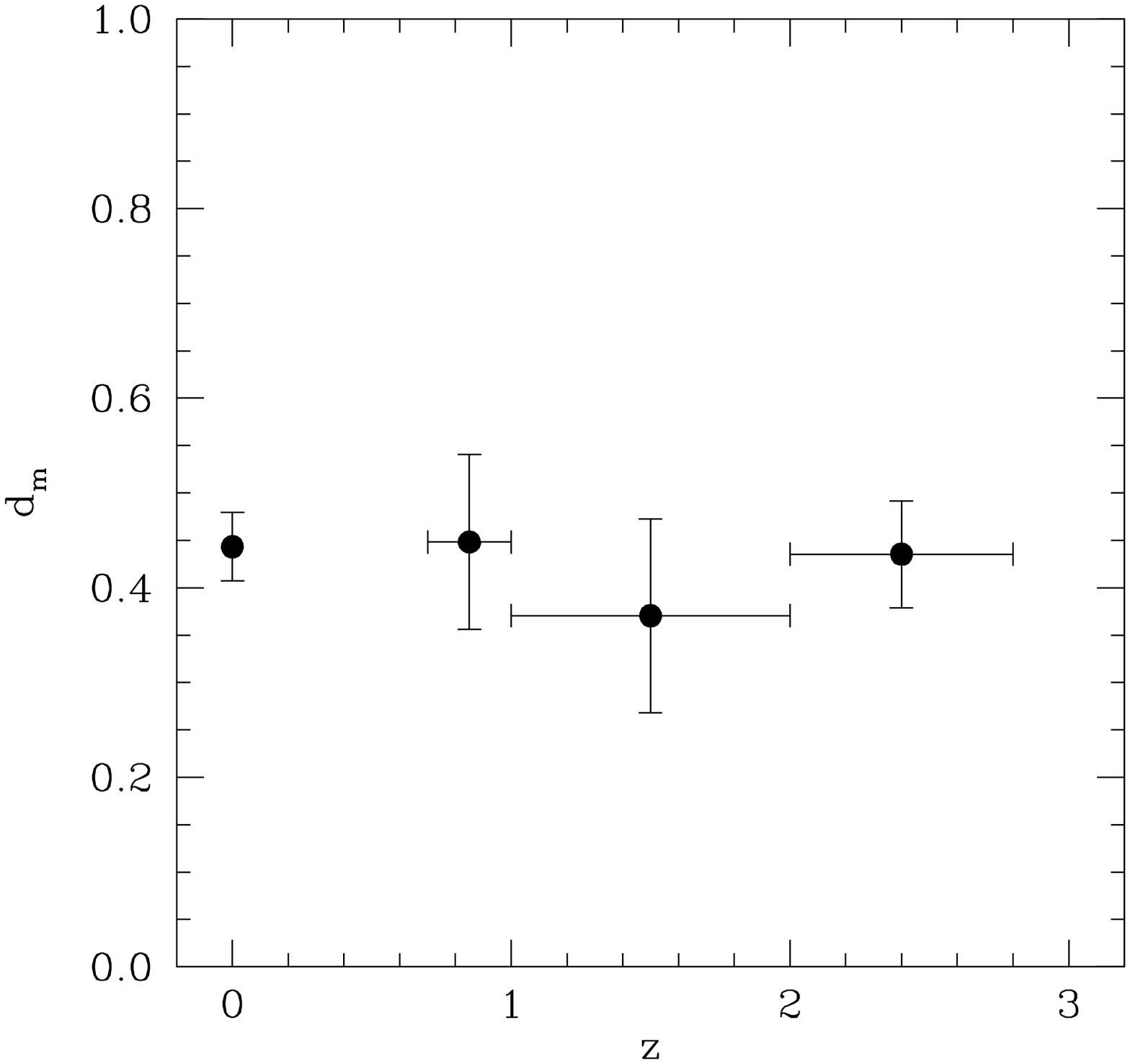,width=3.23in}
\vskip-12pt
F{\sc IG}.~1.---Mean dust-to-metals mass ratio $d_m$ as a
function of redshift.  The data point at $z=0$ is the mean in the
Milky Way, LMC, and SMC (Pei 1992; Luck \& Lambert 1992), while the
others are derived from measurements of the gas-phase depletion of Cr
relative to Zn in damped Ly$\alpha$ systems (Pettini \etal 1997a).
The error bars represent $1\sigma$ uncertainties in the means.
\vskip31pt

The equations presented above for cosmic chemical evolution are quite
general and can be used to study the main baryonic ingredients of
galaxies and their evolution.  The global quantities of interest here
include the comoving densities of stars ($\Omega_s$), interstellar gas
($\Omega_g$), heavy elements ($\Omega_m$), dust ($\Omega_d$), and
baryons ($\Omega_f$) in galaxies, and the comoving rates of star
formation ($\dot\Omega_s$), gas accretion ($\dot\Omega_g$), metal
enrichment ($\dot\Omega_m$), dust production ($\dot\Omega_d$), and
baryon flow ($\dot\Omega_f$), all as functions of redshift.  With
equations~(4)-(6) and straightforward time differentiation and
integration between the comoving density and rate pairs, there are
only two (functional) degrees of freedom in the problem, namely,
inputs of any two independent quantities as functions of redshift
allow all others to be derived.  We choose $\Omega_g(z)$ and
$\dot\Omega_s(z)$ as the two inputs, since both can be traced over a
wide range of redshifts by observations.

\subsection{2.2. Absorption History}

The damped Ly$\alpha$ systems have HI column densities in excess of
$2\times10^{20}~{\rm cm}^{-2}$ and account for more than 80\% of the
neutral gas in the universe (Lanzetta \etal 1991).  The other quasar
absorption-line systems, those with HI column densities less than
$10^{ 20}~{\rm cm}^{ -2}$, probably contain more gas in total than the
damped Ly$\alpha$ systems, but this gas is diffuse and highly ionized.
The damped Ly$\alpha$ systems are usually interpreted as the ancestors
of present-day galaxies, but it is not yet clear whether they are
mainly the progenitors of galactic disks or spheroids, or dwarf
galaxies.  However, the comoving density of HI, the quantity of
interest here, can be computed directly from the statistics of the
absorption along random lines of sight and is therefore independent of
the morphologies, sizes, and other properties of the damped Ly$\alpha$
systems.  The main restriction on our results is that they do not
include any galaxies that consumed their neutral gas before $z\approx
4$, the highest redshift probed systematically by existing surveys of
damped Ly$\alpha$ systems.

\vskip-9pt
\subsubsection{2.2.1. Comoving Density of Gas}
\vskip-3pt

The interstellar content of galaxies consists mainly of hydrogen and
helium (in neutral atomic, ionized, and molecular forms); thus
\vskip-6pt
$$
\Omega_g=\mu\Omega_{\rm H}=\mu(\Omega_{\rm HI}+\Omega_{\rm HII}+
\Omega_{{\rm H}_2}), \eqno(7)
$$
\vskip-6pt\noindent
where $\mu$ is the mean particle mass per H atom, and HI, HII, and
H$_2$ denote neutral atomic, ionized, and molecular hydrogen,
respectively.  The damped Ly$\alpha$ systems are probably mainly
neutral, because their HI column density exceeds $N\gtrsim2\times
10^{20}~{\rm cm}^{-2}$ and their metal absorption lines usually show
low ionization states.  They also appear to have low abundances of
molecules, at least at high redshifts (Levshakov \etal 1992; Ge \&
Bechtold 1997).  As an approximation, we ignore HII and H$_2$ but
include He:
\vskip-6pt
$$
\Omega_g=\mu\Omega_{\rm HI}, \eqno(8)
$$
\vskip-6pt\noindent
with $\mu=1.3$, appropriate for 23\% primordial He by mass from Big
Bang nucleosynthesis (Hogan, Oliver, \& Scully 1997).  While the
neglect of H$_2$ is probably reasonable at high redshifts, it is less
accurate toward $z=0$.  In the local universe, the observed H$_2$ and
HI masses of 154 late-type galaxies (Sa to Im) give an averaged
H$_2$-to-HI mass ratio of roughly 0.8, implying 45\% of H in H$_2$
(Young \& Scoville 1991).  Thus, it appears that equation~(8) is
approximately valid (to within a factor of two) over much or all of
cosmic history.

The comoving density of HI as a function of redshift is given by
$$
\Omega_{\rm HI}(z)={8\pi Gm_{\rm H}\over3cH_0}\int_0^\infty dN N f(N,z),
\eqno(9)
$$
where $m_{\rm H}$ is the mass of the H atom, and $f$ is the
distribution of HI column densities, defined such that $f(N,z)dNdX$ is
the number of absorbers along a random line of sight with HI column
densities between $N$ and $N+dN$ and redshift paths between $X$ and
$X+dX$ [with $dX=(1+z)(1+2q_0z)^{-1/2}dz$ for $\Lambda=0$] (Lanzetta
\etal 1991).  The existing samples of damped Ly$\alpha$ systems are
all derived from optically selected quasars.  As a result of the
obscuration caused by dust within the damped Ly$\alpha$ systems, these
samples are biased to some degree (Fall \& Pei 1993). Thus, we
introduce the observed comoving density of HI
$$
\Omega_{{\rm HI}o}(z)={8\pi Gm_{\rm H}\over3cH_0}\int_0^\infty dN N
f_o(N,z), \eqno(10)
$$
where $f_o$ represents the observed distribution of HI column
densities in damped Ly$\alpha$ systems.  In the limit of an infinitely
large sample of absorbers, the true and observed distributions, $f$
and $f_o$, sample all impact parameters and orientations of galaxies,
and the resulting comoving densities, $\Omega_{\rm HI}$ and
$\Omega_{{\rm HI}o}$, can be derived without any knowledge of the
sizes or shapes of galaxies.

Both $\Omega_{\rm HI}$ and $\Omega_{{\rm HI}o}$ are averages over the
population of damped Ly$\alpha$ absorbers, but the former pertains to
random lines of sight, while the latter pertains to the lines of sight
to optically selected quasars.  Since the missing absorbers in an
optically selected sample tend to be those with the highest $N$, there
can be large differences between $\Omega_{\rm HI}$ and $\Omega_{{\rm
HI}o}$ and this can affect the resulting picture of cosmic chemical
evolution (Pei \& Fall 1995).  We neglect another potential bias, that
due to the magnification of quasars by gravitational lenses associated
with foreground damped Ly$\alpha$ systems, because this appears to be
a relatively minor effect in the samples from which $\Omega_{{\rm
HI}o}$ has been estimated (Le Brun \etal 1997; Perna \etal 1997;
Smette \etal 1997).  Since the dust-to-gas ratios are not known
individually for many damped Ly$\alpha$ systems, the conversion from
$\Omega_{{\rm HI}o}$ to $\Omega_{\rm HI}$ must be made statistically.
For this purpose, we introduce a mean correction factor $Q$, defined
by the relation
\vskip-6pt
$$
\Omega_{\rm HI}(z)=\Omega_{{\rm HI}o}(z)Q(z). \eqno(11)
$$
\vskip-6pt\noindent

\vskip-9pt
\subsubsection{2.2.2. Missing Absorbers}
\vskip-3pt

The relation between the true and observed distributions of HI column
densities can be derived under two idealizations: that the absorbers
are obtained from a magnitude limited sample of bright quasars, and
that all absorbers at a given redshift have the same dust-to-gas
ratio.  In this case, we have (Fall \& Pei 1993)
\vskip-6pt
$$
f(N,z)=f_o(N,z)\exp[\beta\tau(N,z)], \eqno(12)
$$
\vskip-18pt
$$
\tau(N,z)= k_m(z) m_H N \kappa_e[\lambda_e/(1+z)], \eqno(13)
$$
\vskip-18pt
$$
k_m(z)=\Omega_d(z)/\Omega_{\rm HI}(z), \eqno(14)
$$
\vskip-6pt\noindent
where $\beta$ is the power-law index of the bright part of the quasar
luminosity function [i.e., $\phi(L,z)\propto L^{-\beta-1}$], $\tau$ is
the extinction optical depth at $z=0$, $k_m$ is the mean dust-to-HI
mass ratio, $\kappa_e$ is the mass extinction coefficient evaluated at
a wavelength $\lambda=\lambda_e/(1+z)$, and $\lambda_e$ is the
effective wavelength of the limiting magnitude.  Equations (12)-(14)
are also valid for combinations of samples with different limiting
magnitudes, provided these are brighter than $V\lesssim20$, and they
are good approximations when there is a dispersion in the dust-to-HI
ratio, provided this satisfies $\sigma(\log k_m)\lesssim1$.  Both
conditions appear to be true for the existing samples of damped
Ly$\alpha$ systems and background quasars (see \S~2.2.3 and \S~2.2.4).

To derive the correction factor $Q$, we must adopt a specific form for
the observed distribution of HI column densities.  A common
approximation is a power law, $f_o(N,z)\propto N^{-\gamma}$, with
$\gamma\approx1.7$ (Lanzetta \etal 1991).  Unfortunately, this gives a
divergent total mass in equation~(10), unless an upper cutoff in $N$
is imposed.  Thus, following Pei \& Fall (1995), we adopt the gamma
distribution
\vskip-18pt
$$
f_o(N,z)=(f_*/N_*)(N/N_*)^{-\gamma}\exp(-N/N_*),  \eqno(15)
$$
\vskip-6pt\noindent
where $f_*$ and $N_*$ specify the normalization and ``knee'' of the
function.  This parametric form provides a solution to the problem of
diverging mass and a better fit to the data than the power-law form
(Storrie-Lombardi \etal 1996a).  In this case, the true distribution
$f$ from equation (12) is also a gamma distribution
\vskip-18pt
$$
f(N,z)=(f_t/N_t)(N/N_t)^{-\gamma}\exp(-N/N_t),  \eqno(16)
$$
\vskip-6pt\noindent
with $f_t=Q^{\gamma-1\over\gamma-2}f_*$ and $N_t=Q^{1\over2-\gamma}
N_*$.  For $\gamma<2$, the observed and true comoving densities of HI
are simply $\Omega_{{\rm HI}o}=(8\pi Gm_{\rm H}/3cH_0)\Gamma(2-\gamma)
f_* N_*$ and $\Omega_{\rm HI }=(8\pi Gm_{\rm H}/3cH_0)\Gamma(2-\gamma)
f_t N_t$.  Combining equations (9)-(15) and eliminating $N_*$ in favor
of $\gamma$ and $f_*$, we obtain
\vskip-6pt
$$
Q=\beta\tau_*+Q^{\gamma-1\over\gamma-2}, \eqno(17)
$$
\vskip-27pt
$$
\tau_*(\lambda_e,z)\equiv{3cH_0\over8\pi Gf_*\Gamma(2-\gamma)}\Omega_d(z)
\kappa_e[\lambda_e/(1+z)], \eqno(18)
$$
\vskip-3pt\noindent
where $\Gamma$ denotes the standard gamma function.  Equation (17) is
an implicit relation from which $Q$ can be obtained as a function of
$\beta \tau_*$ and $\gamma$.  For $\gamma=1$, we have $Q=1+\beta\tau_*
$ and, for $\gamma=3/2$, we have $Q=[1+(\beta\tau_*/2)^2]^{1/2}+(\beta
\tau_*/2)$.

\vskip-9pt
\subsubsection{2.2.3. Inputs from Damped Ly$\alpha$ Surveys}
\vskip-3pt

The formulae presented above enable us to compute the true $\Omega_g$
from the observed comoving density of HI, $\Omega_{{\rm HI}o}$.  This
relation is
\vskip-7pt
$$
\Omega_g(z)=\mu\Omega_{{\rm HI}o}(z)Q(z),  \eqno(19)
$$
\vskip-7pt\noindent
where $\mu$ accounts for He and $Q$ accounts for the missing damped
Ly$\alpha$ systems in an optically selected sample of quasars.  The
correction factor $Q$ depends on the mean comoving density of dust
$\Omega_d$, the extinction coefficient of grains $\kappa_e$, the
bright-end slope of the quasar luminosity function $\beta$, the
effective wavelength of the quasar sample $\lambda_e$, the slope of
the absorber column density distribution $\gamma$, and the
characteristic sky coverage of the absorbers $f_*$.  Because $Q$ is
linked to $\Omega_d$, our solutions of the equations of cosmic
chemical evolution are obtained iteratively (see \S~3).  In the
following, we adopt $\Omega_{{\rm HI}o}(z)$ at various redshifts from
surveys of the damped Ly$\alpha$ systems and estimate the values of
the parameters $(\beta, \lambda_e,\gamma,f_*)$ appropriate for these
surveys.

The observed HI content in the damped Ly$\alpha$ systems has been
traced over the range of redshift $0.008\le z\le4.7$, in surveys by
Wolfe \etal (1986, 1995), Lanzetta \etal (1991, 1995), and
Storrie-Lombardi \etal (1996b).  We adopt the most recent
determinations of $\Omega_{{\rm HI}o}(z)$, derived by Storrie-Lombardi
\etal (1996b) assuming a particular cosmology with $\Lambda=0$,
$q_0=1/2$, and $H_0=50 $~km~s$^{-1}$~Mpc$^{-1}$.  These estimates,
listed in Table 2 with their errors, are based on 44 damped Ly$\alpha$
systems obtained from a combined sample of 366 optically selected
quasars.  The statistical uncertainty in $\Omega_{{\rm HI}o}(z)$ is
roughly 50\%.  The present value $\Omega_{{\rm HI}o}(0)$ is not
directly determined from absorption in damped Ly$\alpha$ systems, and,
as a result, is not available as input to our models.  Instead, we
appeal to observations of 21~cm emission for the HI content of
galaxies in the local universe (\S~2.2.6).

The existing samples of damped Ly$\alpha$ systems have been obtained
from samples of quasars selected by $V\le18.5$ (Wolfe \etal 1986;
Lanzetta \etal 1991), $B\le18.75$ (Wolfe \etal 1995), $V\lesssim17$
(Lanzetta \etal 1995), and $R\le19.5$ (Storrie-Lombardi \etal 1996a).
Thus, they do not conform to the idealization of a single limiting
magnitude under which equation (12) was derived.  Fortunately, all of
the samples are brighter than $V\approx20$.  In this case, the
luminosity function of quasars is well approximated by a single power
law (Hartwick \& Schade 1990; Pei 1995).  The limiting magnitudes
cancel out and do not appear explicitly in equation (12), except
through the effective wavelength $\lambda_e$.  Therefore, our formulae
for $Q$ are also valid for combinations of samples with different
limiting magnitudes.  The value of $\beta$ can be determined most
simply from the differential counts $A(m)$ of quasars versus apparent
magnitude $m$, because $\phi(L,z)\propto L^{-\beta-1}$ implies
$A(m)\propto 10^{0.4\beta m}$.  Figure~2 shows the observed $B$-band
counts from Hartwick \& Schade (1990).  The line is a weighted,
least-squares fit to the data at $B\le20$, with the $\chi^2$-fitted
value of $\beta$ listed in Table~1.  At $z<3.5$, most of the absorbers
come from the $V$-selected quasars, while at $z\ge3.5$, all of them
come from the $R$-selected quasars.  Accordingly, we adopt $\lambda_e
=5500~${\AA} at $z <3.5$ and $\lambda_e=6500~${\AA} at $z>3.5$.  The
value of $\lambda_e$ at $z=0$ does not enter our calculations but
implies a specific value of $\Omega_{{\rm HI}o}(0)$ for the purpose of
comparison (\S~2.2.6).

The current data for the observed distribution $f_o$ in the damped
Ly$\alpha$ systems are sparse but nevertheless indicate that the
evolution of $f_o$ is strongest for the high-$N$ systems (Wolfe \etal
1995; Storrie-Lombardi \etal 1996a).  In the gamma distribution model,
this is equivalent to varying $N_*$ with respect to $z$, and hence the
evolution of $\Omega_{{\rm HI}o}$ is predominantly due to $N_*$ rather
than $\gamma$ and $f_*$.  Thus, it is a good approximation to regard
$\gamma$ and $f_*$ as constants.  We determine these parameters by
fitting the cumulative form of $f_o$, i.e., the observed number $n_o(
>\!\!N)$ of absorbers with HI column densities greater than $N$.  The
results are shown in Figure~3, where the data (in the stepped line and
the circle) are\break

\vskip20pt
\psfig{file=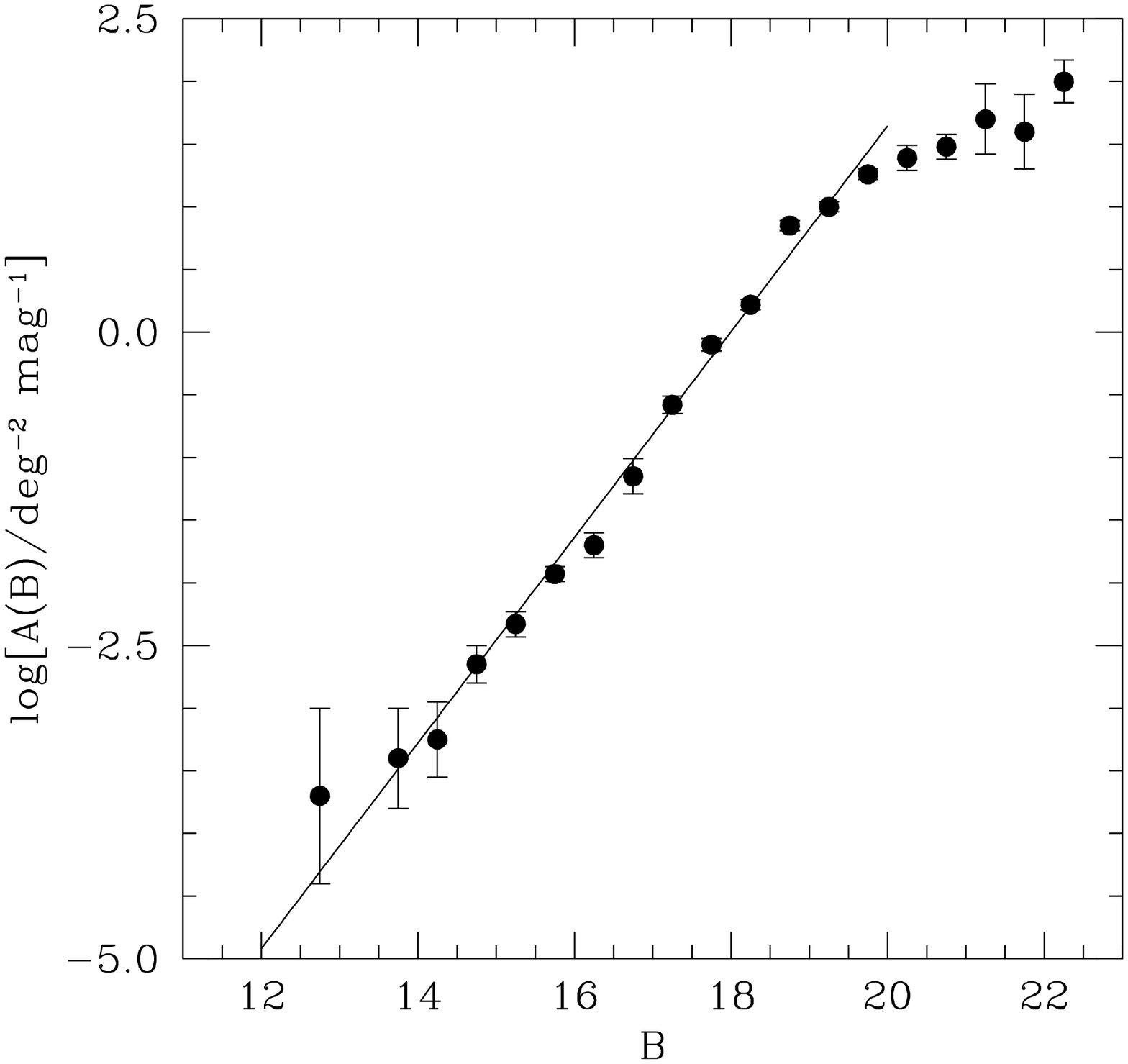,width=3.23in}
\vskip-12pt
F{\sc IG}.~2.---Differential counts of quasars per square degree
versus $B$ magnitude.  The data points were derived by Hartwick \&
Schade (1990) from a combined sample of more than 1000 quasars over
$0\le z\le3.3$.  The line is a weighted, least-squares fit to the
observations at $B\le20$, relevant to the samples of quasars from
which the damped Ly$\alpha$ systems were selected.

\noindent
from Storrie-Lombardi \etal (1996a).  The curve is the
gamma distribution (integrated over the redshift path of all absorbers
in the sample), with the fitted values of $\gamma$ and $f_*$ listed in
Table~1.  The circle at $N=1.6\times10^{17}~{\rm cm}^{-2}$ is the
number of Lyman-limit systems that would be detected along the same
redshift path as in the damped Ly$\alpha$ sample.  We have included it
in our fits to reduce further the uncertainty in $\gamma$ and $f_*$,
especially the former.  The values of $\gamma$ and $f_*$ derived here
are similar to those obtained by Storrie-Lombardi \etal (1996a) from a
maximum likelihood fit.

\vskip-10pt
\subsubsection{2.2.4. Dispersion in the Dust-to-Gas Ratio}
\vskip-3pt

The main idealization in our derivation of the mean correction $Q$ is
the assumption that all absorbers at a given redshift have the same
dust-to-gas ratio.  This gives a simple relation between the true and
observed distributions in equation (12) so that $Q$ can be evaluated
using $f_o$ (or $f$).  It also links the dust-to-HI ratio $k_m$ to the
comoving density of dust in equation (14) so that $Q$ is coupled
self-consistently to the equations of chemical evolution.  The damped
Ly$\alpha$ systems do show a significant dispersion in their
dust-to-gas ratios, as indicated by the observed Cr/Zn ratios (Pettini
\etal 1997a).  A dispersion in the dust-to-gas ratio would\break

\vskip19.5pt
\psfig{file=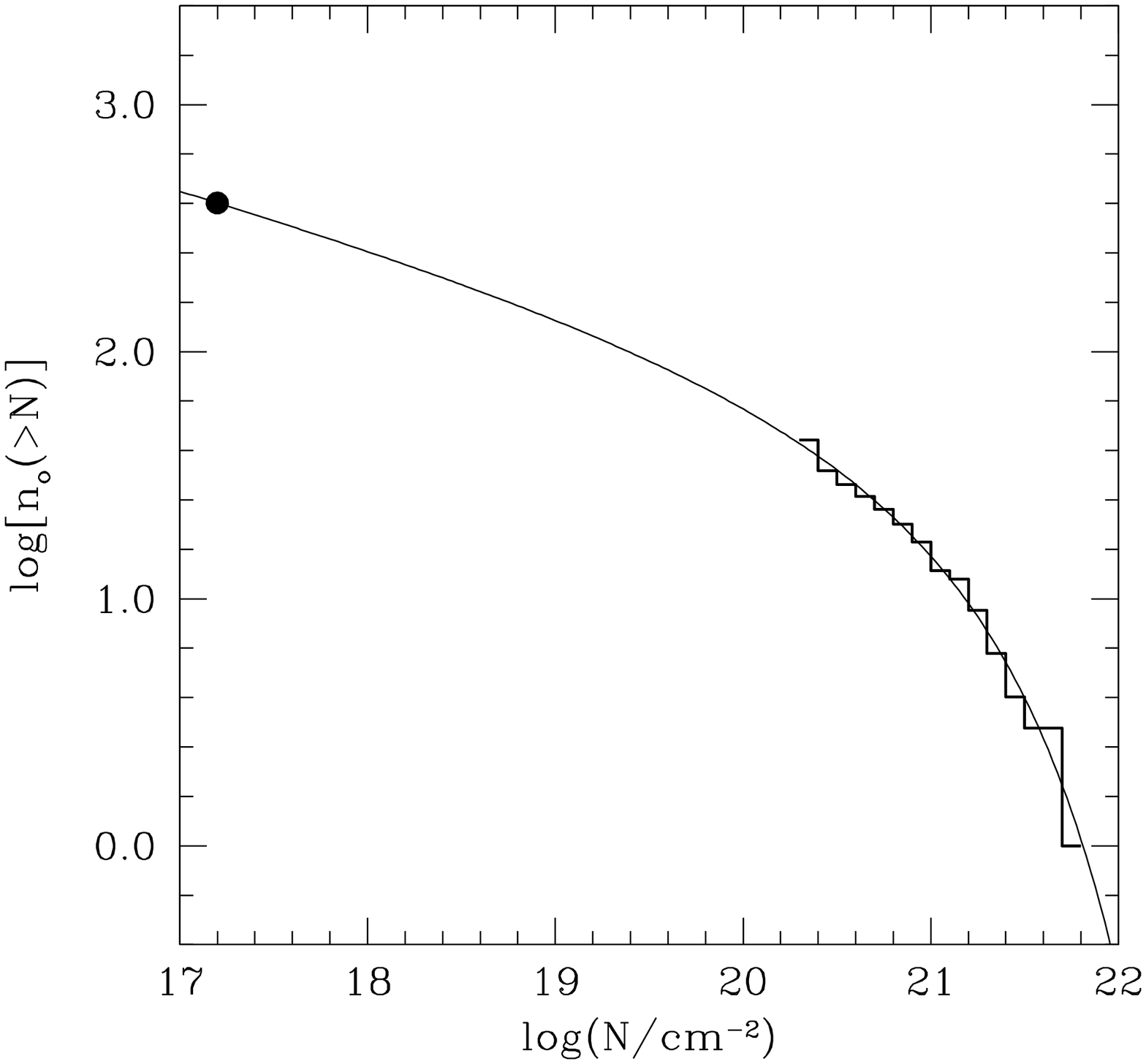,width=3.23in}
\vskip-12pt
F{\sc IG}.~3.---Cumulative number of absorption systems with HI
column densities greater than $N$.  The stepped line represents the
data from existing surveys of damped Ly$\alpha$ systems (with a total
redshift path $\Delta X=421$), while the point is the mean number of
Lyman limit systems that would be detected along the same redshift
path (Storrie-Lombardi \etal 1996b).  The curve shows the fit of the
cumulative form of the gamma distribution.

\noindent
introduce
additional terms of $N$ in the relation between $f$ and $f_o$, rather
than the simple relation in equation (12) (see the Appendix of Fall \&
Pei 1993).  Nearby galaxies have $\sigma(\log k_m)\approx0.4-0.5$ (Pei
1992) and the observed damped Ly$\alpha$ systems have $\sigma(\log
k_m)\approx0.6$ at ${0.7<z<2}$ and $0.6$ at $2<z<2.8$ (Pettini \etal
1997a).  If the true dispersion were not much larger than the observed
one, the additional terms in the relation between $f$ and $f_o$ would
be negligible (less than roughly 20\% at $N=10^{21}~{\rm cm}^{-2}$ for
$k_m=0.1~Z_\odot$).  Thus, the mean correction $Q$ derived here is
not sensitive to a modest dispersion in the dust-to-gas ratio [i.e.,
$\sigma(\log k_m)\lesssim1$].

A dispersion in the dust-to-gas ratio could, however, affect the
observed mean dust-to-gas ratio and metallicity in damped Ly$\alpha$
absorbers.  The observed mean dust-to-gas ratio could be
underestimated, because the absorbers with highest dust-to-gas ratios
at fixed HI column density would be preferentially missing from
optically selected quasars.  Using the formulae of Fall \& Pei (1993),
we estimate that the observed mean dust-to-gas ratio would be
$30\%-50\%$ lower than the true mean for $\sigma(\log
k_m)\approx0.4-0.6$ (at $k_m=0.1~Z_\odot$).  If the heavy elements in
individual galaxies were correlated with the dust, the observed mean
metallicity would also be underestimated by a similar amount.  These
biases, while not negligible, are comparable to the observational
uncertainties in the derived mean dust-to-gas ratios and metallicities
in the current samples of damped Ly$\alpha$ systems.  We emphasize
that our models of cosmic chemical evolution do not use as input the
observed comoving densities of dust and metals, and consequently they
are not affected by a dispersion in the dust-to-gas ratio.  The only
complication is that, since no corrections are made, this could affect
the comparison of our model outputs with the observed mean dust-to-gas
ratio and metallicity in the damped Ly$\alpha$ systems.

\vskip-9pt
\subsubsection{2.2.5. Optical Properties of Grains}
\vskip-3pt

The optical properties of interstellar dust grains at high redshifts
are largely unknown.  We must therefore appeal to the properties of
galaxies at the present epoch.  The Milky Way, LMC, and SMC are three
nearby galaxies with well observed interstellar extinction curves.  To
derive self-consistently the wavelength dependence of the albedo
(relevant for the absorption of starlight by dust discussed in
\S~2.3), we adopt the Draine \& Lee (1984) grain model but with the
proportions of graphite and silicates adjusted so as to fit the mean
empirical extinction curves in the Milky Way, LMC, and SMC (Pei 1992).
Figure 4 shows the resulting opacities, $\kappa_e$ (mass extinction
coefficient in the upper panel) and $\kappa_a$ (mass absorption
coefficient in the lower panel), as functions of wavelength in the
three galaxies.  The range in $\kappa_e$ between the Milky Way, LMC,
and SMC is roughly 30\%, whereas the range in $\kappa_a$ can be as
large as a factor of two, especially at optical wavelengths.  The LMC
and SMC have lower abundances of metals and dust than the Milky Way
and may be more representative of galaxies at high redshifts.  There
is some evidence for this from the weak or absent 2200~{\AA}
``absorption'' feature in the damped Ly$\alpha$ systems (Pei \etal
1991).  In our calculations, we adopt the LMC-type dust, representing
an intermediate case between the Galactic-type and SMC-type dust.  We
will later discuss the effects of different $\kappa_a$ on our results
(\S~4.2).

\vskip16pt
\psfig{file=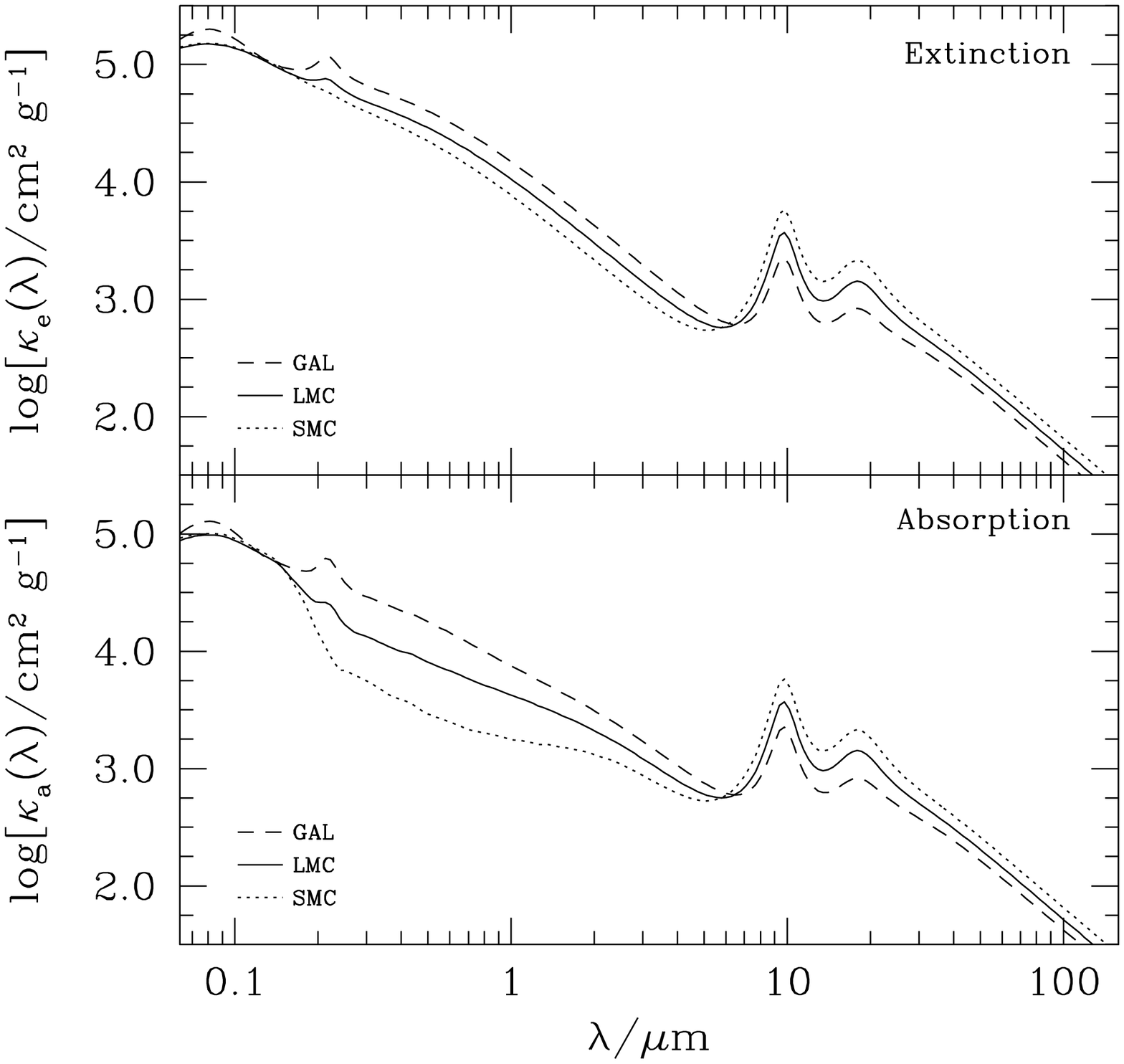,width=3.23in}
\vskip-12pt
F{\sc IG}.~4.---Extinction and absorption coefficients (per unit
mass) of grains in the Milky Way, LMC, and SMC.  These are based on
the Draine \& Lee (1984) model but with the proportions of graphite
and silicates adjusted to fit the observed mean extinction curves in
the three galaxies (Pei 1992).
\vskip12pt

\vskip-9pt
\subsubsection{2.2.6. Neutral Gas in the Local Universe}
\vskip-3pt

The existing samples of damped Ly$\alpha$ systems contain only a few
absorbers at $z\lesssim2$, causing nearly 60\% uncertainty in the
corresponding estimates of $\Omega_{{\rm HI}o}(z)$ in Table~2.  The
value of $\Omega_{{\rm HI}o}$ at $z=0$, if extrapolated from these
entries, is even more uncertain.  To minimize this uncertainty, we
note that averages over the HI content of galaxies in the local
universe give $\Omega_{\rm HI}(0)=(3.8\pm0.9)\times10^{-4}$ (Fall \&
Pei 1993; Rao \& Briggs 1993; Zwaan \etal 1997).  This estimate of
$\Omega_{\rm HI}(0)$ is based on 21~cm emission from galaxies, and
unlike $\Omega_{ {\rm HI}o}(0)$, it is not affected by dust
obscuration.  Thus, we adopt this to fix the present true comoving
density of gas in our models, listed in Table~2 as the first entry
under $\Omega_g(z)$.  Galaxies at $z=0$ contain dust and hence cause
some difference between $\Omega_{\rm HI}(0)$ and $\Omega_{{\rm
HI}o}(0)$.  The value of $\Omega_{\rm HI}(0)$ adopted here implies
$\Omega_{{\rm HI}o}(0)=(2.2-2.7)\times10^{-4}$ in our models,
consistent with the range of $\Omega_{{\rm HI}o}(0)=(2.1-3.0)\times
10^{-4}$ that would be extrapolated from the data in the damped
Ly$\alpha$ systems [with a form $\log\Omega_{{\rm HI}o}(z)\propto z$
or $\log\Omega_{{\rm HI}o} (z)\propto\log(1+z)$ at $z\lesssim2$].

\subsection {2.3. Emission History}

We now turn to the second input to our models, $\dot\Omega_s(z)$.
Deep optical imaging surveys probe the rest-frame ultraviolet
radiation of high-redshift galaxies and the rest-frame optical
radiation of low-redshift galaxies.  When combined with redshift
estimates, they trace the stellar emission history of galaxies.  The
ultraviolet light is produced mainly by short-lived, massive (O and
early B) stars and can be used to infer the instantaneous star
formation rate, nearly independent of the prior star formation
history.  However, ultraviolet radiation may be strongly absorbed by
dust, and the star formation rate may be severely underestimated if no
corrections are made.  Since the absorbed starlight is reradiated by
dust, the recent measurements of the cosmic infrared background by the
DIRBE and FIRAS provide a crucial constraint, which allows us to
correct for any radiative energy that is missing in the optical data.
The infrared background is the cumulative result of emission at all
redshifts.  When combined with the chemical equations, the redshift
information is then decoded from the integrated background constraint.
This section provides a self-consistent treatment of the absorption
and reradiation by dust in order to obtain unbiased estimates of the
global rates of star formation.

\vskip-9pt
\subsubsection{2.3.1. Emissivity and Background Intensity}
\vskip-3pt

We are interested in the global evolution of the radiation from stars
and dust in galaxies.  This evolution can be described naturally in
terms of the comoving emissivity $E_\nu(z)$ as a function of redshift,
defined here as the power radiated per unit frequency per unit
comoving volume at the redshift $z$.  When computing $E_\nu$, we
include only photons that have escaped from galaxies and neglect the
absorption of infrared photons.  This emissivity is then a sum of the
direct (but attenuated) starlight and the reradiated starlight from
dust in galaxies:
\vskip-6pt
$$
E_\nu=(1-A_\nu)E_{s\nu}+E_{d\nu}, \eqno(20)
$$
\vskip-6pt\noindent
where $A_\nu$ is the mean fraction of photons absorbed by dust in the
interstellar medium, and $E_{s\nu}$ and $E_{d\nu}$ are the stellar and
dust emissivities, respectively.  It is clear that $E_{s\nu}$
dominates at ultraviolet, optical, and near-infrared wavelengths,
while $E_{d\nu}$ dominates at mid-infrared, far-infrared, and
submillimeter wavelengths.  The cosmic background intensity $J_\nu$ at
$z=0$, defined as the power received per unit frequency per unit area
of detector per unit solid angle of sky, is given by an integral of
$E_\nu(z)$ over $z$:
\vskip-9pt
$$
J_\nu={c\over4\pi}\int^{\infty}_0 dz E_{\nu(1+z)}\biggl|{dt\over
dz}\biggr|. \eqno(21)
$$
We have ignored any external absorption along the line of sight
between the sources of radiation and the observer, because all
ionizing photons are assumed to be absorbed locally within the
galaxies in which they were produced, and because relatively few
non-ionizing photons are absorbed by the dust in intervening galaxies.
We have checked that, when external absorption by dust is included in
our models, only 3\% of the observed $V$-band radiation is absorbed
along the line of sight from $z=5$ to $z=0$.

\vskip-9pt
\subsubsection{2.3.2. Stellar Emissivity}
\vskip-3pt

The intrinsic stellar emissivity (before absorption) can be expressed
in terms of the comoving rate of star formation $\dot\Omega_*$:
\vskip-5pt
$$
E_{s\nu}=\rho_c\int_0^{t} dt' F_\nu(t-t')\dot\Omega_*(t'),
\eqno(22)
$$
\vskip-4pt\noindent
where $F_\nu(\Delta t)$ is the stellar population spectrum, defined as
the power radiated per unit frequency per unit initial mass by a
generation of stars with an age $\Delta t$.  The total rate of star
formation $\dot\Omega_*$ differs from the net rate $\dot\Omega_s$
introduced in \S~2.1 because some of the stellar material is ejected
back to the interstellar medium by evolving and dying stars; thus
\vskip-6pt
$$
\dot\Omega_*=(1-R)^{-1}\dot\Omega_s, \eqno(23)
$$
\vskip-6pt\noindent
where $R$ is the IMF-averaged returned fraction.  The observed stellar
emissivity (after absorption), denoted by $E_{\nu o}$, is then given
by
\vskip-6pt
$$
E_{\nu o}\equiv(1-A_\nu)E_{s\nu}=(1-A_\nu)C_\nu\dot\Omega_s, \eqno(24)
$$
\vskip-6pt\noindent
where
\vskip-6pt
$$
C_\nu\equiv{\rho_c\over\dot\Omega_s(t)}
\int_0^{t} dt' F_\nu(t-t'){\dot\Omega_s(t')\over1-R(t')} \eqno(25)
$$
is a conversion factor between the intrinsic stellar emissivity and
star formation rate, which depends in general on the past history of
$\dot \Omega_s$.  At ultraviolet wavelengths, however, $C_\nu$ remains
roughly constant in time, nearly independent of the past history of
$\dot \Omega_s$, and hence $E_{\nu o}(z)$ scales directly with $\dot
\Omega_s(z)$ in the absence of dust absorption.

We adopt the latest version of the Bruzual-Charlot population
synthesis code to compute $F_\nu$ and \hskip-0.2pt $R$ (Bruzual \&
Charlot 1998).
The stellar IMF is assumed to be a power law, $\phi(m)\propto
m^{-(1+x)}$, with $x=1.35$ (a Salpeter IMF) and lower and upper mass
cutoffs at $0.1~M_\odot$ and $100~M_\odot$.  We further assume that
all hydrogen ionizing photons are absorbed locally in the interstellar
medium, where 68\% of them are converted to Ly$\alpha$ photons (the
fraction appropriate for case B recombination in gas at $10^4$~K).
The Ly$\alpha$ photons are assumed to be absorbed by dust (as a result
of resonant scattering by HI), and the remaining energy is radiated
uniformly in wavelength between 3000 and 7000~{\AA} (a range that
includes most of the relevant emission lines).  This treatment of
ionizing radiation is consistent with ultraviolet observations of
local starburst galaxies (Leitherer \etal 1995).  In our calculations,
templates of $F_\nu$ and $R$ are pre-computed for metallicities
$Z/Z_\odot = 2.5$, 1.0, 0.4, 0.2, and 0.02.  The metallicity
dependence of the spectral evolution is included self-consistently by
interpolating the templates according to the solution for $Z$ in our
models of cosmic chemical evolution.  Examples of these templates are
shown in Figure~5.  We will later discuss\break
\noindent
the effects of changes in
IMF on our results (\S~4.2).

\vskip16pt
\psfig{file=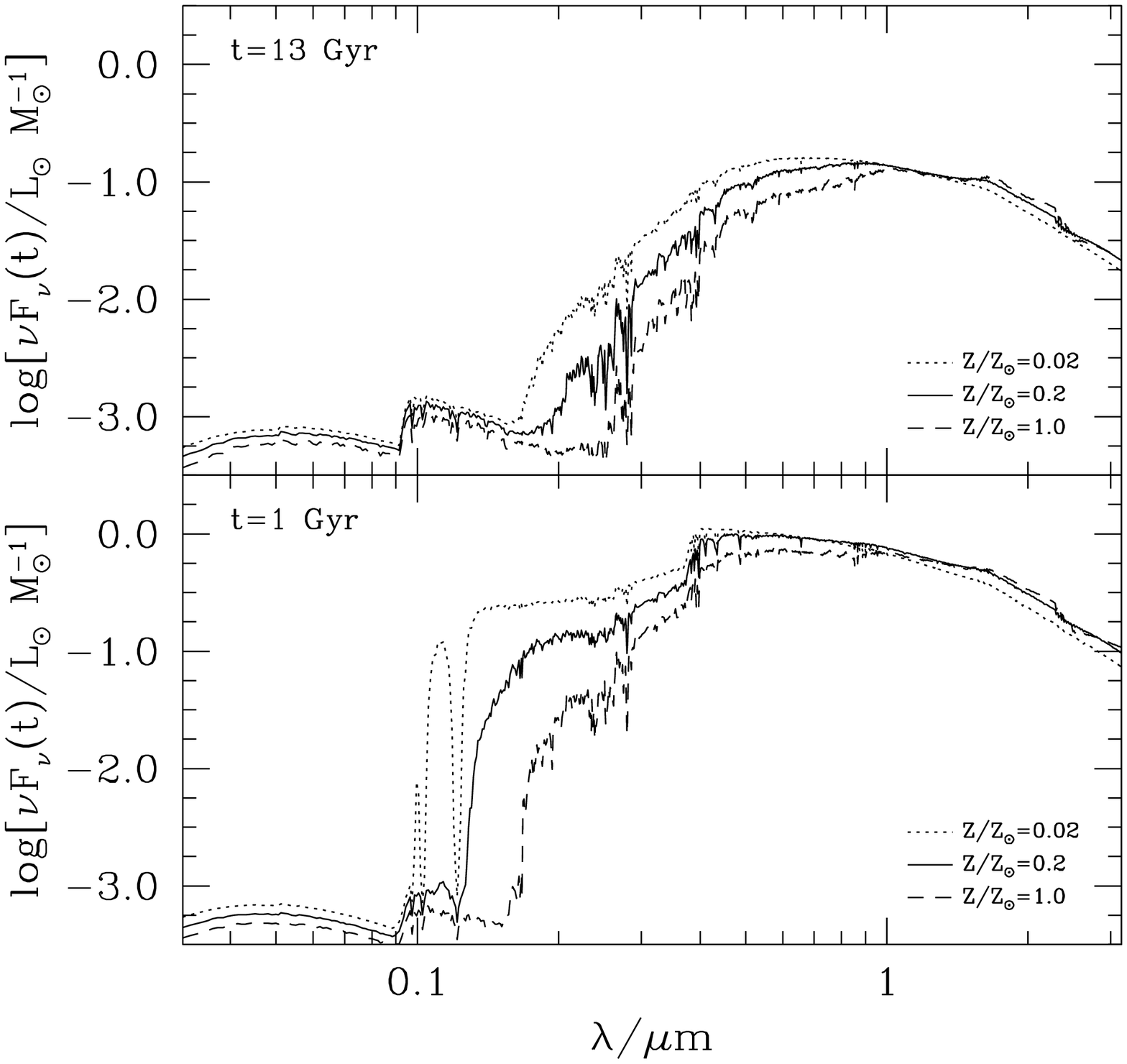,width=3.23in}
\vskip-12pt
F{\sc IG}.~5.---Spectra of stellar populations with different
ages and metallicities.  These are from the latest version of the
Bruzual-Charlot (1998) population synthesis code with a Salpeter IMF
truncated at 0.1 and 100~$M_\odot$.
\vskip10pt

\def\ta{{\tau_\nu}}\def\tp{{\tau^{\prime}_\nu}}

\vskip-9pt
\subsubsection{2.3.3. Absorption of Starlight by Dust}
\vskip-3pt

We now consider the mean fraction $A_\nu$ of photons absorbed by the
interstellar dust in galaxies.  This absorption is likely to occur on
various scales of the dust distribution, ranging from small clumps of
dust shrouding or in the vicinity of individual stars to large clouds
in the interstellar medium.  To account for all scales, it is
appropriate first to consider the absorption along a ray passing
through a galaxy and then to average over all galaxies and all impact
parameters and inclination angles to derive the mean absorption.
Along such a ray with fixed absorption optical depth $\ta$, the
emerging intensity is given by
$$
I_\nu(\ta)=\int_0^\ta d\tp S_\nu(\tp)\exp[-(\ta-\tp)], \eqno(26)
$$
\vskip-2pt\noindent
where $S_\nu$ is the source function along the ray.  In the absence of
absorption, the corresponding intensity would be
$$
I^0_\nu(\ta)=\int_0^{\ta} d\tp S_\nu(\tp).  \eqno(27)
$$
\vskip-2pt\noindent
We have ignored the influence of scattering on absorption so that
$S_\nu$ does not contain a contribution from $I_\nu$.  To perform the
average over galaxies, we define $g(\ta,z)d\ta dX$ as the number of
galaxies along a random ray with optical depths between $\ta$ and
$\ta+d\ta$ and redshift paths between $X$ and $X+dX$ when all galaxies
and all positions and directions within them are considered.  The
redshift path, defined by $dX=(1+z)(1+2q_0z)^{-1/2 }$ for $\Lambda=0$,
is included in the definition of $g(\ta,z)$ so that an integral of
$\ta g(\ta,z)$ over all $\ta$ gives the comoving density of dust
$\Omega_d$ (see eq.~36).  With this definition, we can express the
mean fraction of photons absorbed by dust in the form
$$
A_\nu(z)={\int_0^\infty d\ta g(\ta,z)[I^0_\nu(\ta)-I_\nu(\ta)]
\over\int_0^\infty d\ta g(\ta,z) I^0_\nu(\ta)}.  \eqno(28)
$$
\vskip4pt

The distribution function $g(\ta,z)$ could in principle be modeled by
specifying in detail the three-dimensional dust distribution in
galaxies of different types, sizes, etc.  This would, however, require
a large number of assumptions.  Instead, we appeal to damped
Ly$\alpha$ surveys to specify $g(\ta,z)$ as a function of both $\ta$
and $z$.  In the limit of infinitely many lines of sight, these
surveys sample correctly all impact parameters and inclination angles
of galaxies at different redshifts.  However, the lines of sight in
quasar absorption-line surveys have finite resolution, due to the
``beam size'' of quasar emission regions.  Although the exact sizes of
quasar emission regions are uncertain, the optical and X-ray continuum
regions are often assumed to be much smaller than $0.1-1$~pc, which is
roughly the size of the broad-line emission region.  Thus, even in the
limit of infinitely many lines of sight, the resulting distribution
function from damped Ly$\alpha$ surveys would not fully sample any
small-scale clumpiness of the dust distribution, probably on scales
much less than 1~pc but certainly on scales of clumps in the vicinity
of individual stars (including circumstellar dust).  This
distribution, denoted by $h (\ta,z)$, has the same definition as
$g(\ta,z)$, except that the former is an unsmoothed version of the
latter.

The relation between $g$ and $h$ can be written as
$$
g(\tp,z)=\int_0^\infty d\ta p(\tp|\ta) h(\ta,z), \eqno(29)
$$
where $p(\tp|\ta)d\tp$ is the probability of finding dust clumps with
optical depths between $\tp$ and $\tp+d\tp$ within a small but finite
beam given that the beam has a mean optical depth $\ta$.  Clearly,
this conditional probability must satisfy the following relations
\vskip-2pt
$$
\int_0^\infty d\tp p(\tp|\ta) = 1, \eqno(30)
$$
\vskip-8pt
$$
\int_0^\infty d\tp \tp p(\tp|\ta) = \ta. \eqno(31)
$$
\vskip-2pt\noindent
If the small-scale clumpiness of the dust distribution within galaxies
were ignored, i.e., $p(\tp|\ta)= \delta(\tp-\ta)$, we would have
$g(\tp,z)= h(\tp,z)$.  Thus, the probability distribution $p$
introduces the additional small-scale clumpiness of the dust
distribution specified by $g$ but not sampled by $h$.

We relate the distribution $h$ of optical depths to the distribution
$f$ of HI column densities in the damped Ly$\alpha$ systems (Fall
\etal 1996).  The absorption optical depth $\ta$ and HI column density
$N$ are related by
\vskip-6pt
$$
\ta=k_m m_{\rm H} N \kappa_a(\nu), \eqno(32)
$$
\vskip-6pt\noindent
where $\kappa_a$ is the mass absorption coefficient of grains and
$k_m$ is the dust-to-HI mass ratio (eq.~14).  Again, we assume that
$k_m$ is only a function of redshift (see \S~2.2.2 and \S~2.2.4), so
that the distributions of $\ta$ and $N$ are simply related by
$h(\ta,z)d\ta=f(N, z)dN$.  From equation (16), we then obtain
\vskip-16pt
$$
h(\ta,z)=(f_t/\tau_t)(\ta/\tau_t)^{-\gamma}\exp(-\ta/\tau_t),
\eqno(33)
$$
\vskip-28pt
$$
\tau_t(\nu,z)\equiv[3cH_0/8\pi Gf_t\Gamma(2-\gamma)]\Omega_d(z)\kappa_a
(\nu), \eqno(34)
$$
\vskip-4pt\noindent
where $\tau_t$ is the characteristic optical depth of the absorbers.
Given the comoving density of dust $\Omega_d$, the absorption opacity
of grains $\kappa_a$, and the parameters $f_t$ (closely related to
$f_*$) and $\gamma$ discussed in \S~2.2.4, the distribution function
$h$ is completely specified by equations (33) and (34).  It is known
that the distribution of gas in the interstellar medium is not a
perfect tracer of the dust distribution, especially on small scales.
In this case, the distribution function $h$, when related to $f$ by a
constant $k_m$, would also undersample some of the small-scale
clumpiness of the dust in galaxies.

The probability distribution $p$ specifies the small-scale clumpiness
of dust in the interstellar medium.  The dust could be in a variety of
forms (e.g., shells, spheres, and other complicated shapes) and in a
variety of places (e.g., surrounding, in the vicinity of, and between
stars).  Instead of specifying the geometry and location of such
clumps relative to the sources of radiation, we adopt a statistical
approach to derive $p$.  Since we are interested in the mean quantity
$A_\nu$ in galaxies, we assume as an idealization that all clumps at a
given redshift have the same optical depth $\tau_c$ (but not
necessarily the same densities, shapes, and sizes), and that such
clumps are randomly distributed within the interstellar medium.  The
probability of intercepting such clumps in a small but finite beam
passing through galaxies with a given mean optical depth $\ta$ is then
given by the usual Poisson distribution:
$$
p(\tp|\ta)=\sum_{n=0}^{\infty} {1\over n!}(\ta/\tau_c)^n\exp(-\ta/
\tau_c) \delta(\tp-n\tau_c). \eqno(35)
$$
This is similar to the clump model developed by Nata \& Panagia
(1984).  It is easy to check that this probability distribution
satisfies equations (30)-(31) and that the small-scale clumpiness of
dust introduced by $p$ does not affect the total amount of dust, i.e.,
\vskip-2pt
$$
\Omega_d(z)={8\pi G\over3cH_0\kappa_a}\int_0^\infty d\ta\ta g(\ta,z)
$$
\vskip-8pt
$$
\hskip30pt={8\pi G\over3cH_0\kappa_a}\int_0^\infty d\ta\ta h(\ta,z). 
\eqno(36)
$$
\vskip-2pt\noindent
Thus, regardless of how clumpy the dust is, the mean comoving density
of dust $\Omega_d$ computed from $h$ is the same as the one computed
from $g$.  This is important because the mean correction factor $Q$
derived in \S~2.2.2 is also the same regardless of whether $h$ or $g$
is used.

To derive the mean fraction of photons absorbed by dust $A_\nu$, we
further assume that the source function $S_\nu$ is constant.
Combining equations (28), (29), and (33)-(35), we obtain
$$
A_\nu=1-{1\over(\gamma-1)\tau_t}\biggl\{\biggl[1+{\tau_t\over\tau_c}
(1-e^{-\tau_c})\biggr]^{\gamma-1}-1\biggr\}. \eqno(37)
$$
For $\tau_c=0$ and $\gamma=1$, this reduces to the formula derived by
Fall \etal (1996).  In our new formula, the absorption of starlight by
dust depends on the comoving density of dust $\Omega_d$ through
$\tau_t$ and the small-scale clumpiness of the dust through $\tau_c$.
The dependence of $A_\nu$ on $\tau_t$ arises from the absorption of
photons traveling through the whole interstellar medium when
small-scale dust clumps are unresolved, while the dependence on
$\tau_c$ includes the additional absorption by such clumps.  In the
limit of $\tau_t\gg\tau_c$, the absorption is dominated by the dust
distribution on scales resolved by quasar absorption-line surveys,
whereas, in the limit of $\tau_t\ll \tau_c$, the absorption is
dominated by the small-scale clumpiness of the dust.  For fixed values
of $\tau_t$ and $\tau_c$, $A_\nu$ depends only mildly on the
assumption of a constant source function.  For example, for fixed
$\tau_t=0.1$ (1.0) and $\tau_c=0$, $A_\nu$ increases by 1.4\% (8\%) if
the sources are all in the middle of the dust (e.g., stars embedded in
clumps) and decreases by 2.6\% (17\%) if the dust is in the middle of
the sources (e.g., clumps between stars), both relative to a constant
$S_\nu$.  Therefore, the assumption of a constant source function is
made here for simplicity and should have only a minor effect on
$A_\nu$.

The parameter $\tau_c$ is by definition an absorption optical depth,
and hence its dependence on frequency is through the absorption
coefficient $\kappa_a$.  It may also have some dependence on redshift
as a result of the chemical enrichment and other processes within
galaxies.  To allow for a range of possibilities, we parameterize this
evolution by
$$
\tau_c(\nu,z)=\tau_{cV}(0)\biggl[{\kappa_a(\nu)\over\kappa_a(V)}\biggr]
\biggl[{\Omega_d(z)\over\Omega_d(0)}\biggr]^{\epsilon}, \eqno(38)
$$
where $\tau_{cV}(0)$ is the visual optical depth of dust clumps at
$z=0$ and $\epsilon$ is the power-law index that links $\tau_c$ to
$\Omega_d$.  For $\epsilon=1$, $\tau_c$ would evolve with redshift the
same as $\Omega_d$.  This might be the case if the metals ejected by
stars (including those locked into dust grains) were instantaneously
mixed in the interstellar medium, so that $\tau_c$ tracked the global
chemical enrichment in galaxies.  For $\epsilon=0$, $\tau_c$ would
remain constant in time.  This might be the case if the timescale for
mixing were longer than the Hubble time, so that most of the metals
remained in the vicinity of stars with little or no evolution.  Given
these arguments, one might expect that a plausible value of $\epsilon$
is somewhere between the cases of instantaneous mixing ($\epsilon=1$)
and the incomplete mixing ($\epsilon=0$).  We treat both $\tau_{cV}$
and $\epsilon$ as adjustable parameters in our models.

\vskip-9pt
\subsubsection{2.3.4. Dust Emissivity}
\vskip-3pt

The stellar radiation absorbed by dust is reradiated thermally in the
infrared.  This emission covers a wide range of wavelengths,
reflecting a wide range of grain temperatures.  The far-infrared
emission is dominated by large amounts of cold dust heated by a smooth
field of radiation from many stars.  The mid-infrared emission is
likely dominated by small amounts of warm and hot dust near stars,
including polycyclic aromatic hydrocarbons or very small grains
producing some non-equilibrium emission.  We write the dust emissivity
as a sum of thermal emission at different temperatures:
\vskip-3pt
$$
E_{d\nu}=4\pi\rho_c\Omega_d\kappa_a(\nu)\int_0^\infty dT B_\nu(T)\eta(T)
$$
\vskip-15pt
$$
\hskip-39pt=4\pi\rho_c\Omega_d\kappa_a(\nu)\langle B_\nu\rangle, \eqno(39)
$$
\vskip-7.5pt\noindent
where $\eta(T)dT$ is the fraction of grains with temperatures between
$T$ and $T+dT$ and $\langle B_\nu\rangle$ denotes the average of
blackbodies $B_\nu(T)$ over the temperature distribution $\eta(T)$.
If all grains had the same temperature $T_d$, i.e., $\eta(T)=\delta(T-
T_d) $, $\langle B_\nu\rangle$ would reduce to the spectrum of a
single blackbody.  The temperature distribution $\eta$ introduces the
warm and hot components of the interstellar dust to account for the
mid-infrared emission.

The exact form of the dust temperature distribution is unknown.  To
specify it fully would require some detailed assumptions about
variations of the radiation field, grain spatial density, and optical
properties (including the size distribution and the grain composition)
throughout galaxies.  Since we are more interested in the far-infrared
background, where accurate measurements exist, and less interested in
the mid-infrared background, where only observational limits exist, we
model the dust temperature distribution simply by a power law
$$
\eta(T)={\alpha-1\over T_c}\biggl({T\over T_c}\biggr)^{-\alpha}
\theta(T-T_c). \eqno(40)
$$
Here, the proportionality is fixed by the requirement that $\eta$ must
be normalized to unity.  We have included a truncation factor $\theta
(T-T_c)$ to exclude $T<T_c$, because all grains, even those far from
stars, will be heated to finite temperatures by the diffuse radiation
field of the whole stellar population (or by the cosmic microwave
background radiation if this exceeds the diffuse starlight in
galaxies).  Since the total emission must be finite, the dust
temperature index in equation (40) must satisfy $\alpha>5+n$ for
$\kappa_a(\nu)\propto\nu^n$.  Indeed, the observed mean infrared
colors of nearby galaxies can be fitted with a value of $\alpha$ that
satisfies this requirement (\S~2.3.5).  For simplicity, we have not
imposed any upper cutoff temperature on $\eta$, since this has little
effect on our results when $\alpha$ is large.  Such a steep slope is
the consequence of a very small fraction of mass in warm and hot dust
in the interstellar medium.

The lower cutoff temperature $T_c$ in equation (40) is closely related
to a mean temperature of dust $T_d$, defined here by the first moment
of $T$ over $\eta(T)$:
\vskip-3pt
$$
T_d\equiv\int_0^\infty dT T \eta(T)=\biggl({\alpha-1\over\alpha-2}
\biggr) T_c. \eqno(41)
$$
\vskip-3pt\noindent
Given the dust temperature index $\alpha$, the averaged blackbody
spectrum $\langle B_\nu\rangle$ is then a unique function of the mean
dust temperature $T_d$.  We determine the redshift dependence of this
mean temperature by the usual condition of energy balance
\vskip-15pt
$$
4\pi\rho_c\Omega_d\int_0^\infty d\nu\kappa_a(\nu)[\langle B_\nu\rangle
(T_d)-B_\nu(T_{\rm CMB})]
$$
$$
=\int_0^\infty d\nu A_\nu E_{s\nu}, \eqno(42)
$$
\vskip-4.5pt\noindent
where $T_{\rm CMB}=2.728(1+z)~$K is the temperature of the cosmic
microwave background radiation (Fixsen \etal 1996).  The inclusion of
the cosmic microwave background has little effect on $T_d$ at $z
\lesssim 7$ in our models.  We have ignored other sources of radiation
(primarily active galactic nuclei) in the heating of interstellar
dust.  The blue emissivity of quasars, estimated from the observed
luminosity function (Pei 1995), is less than a few percent of the blue
emissivity of galaxies at all redshifts in our models.  The
far-infrared emissivity of quasars, however, is not yet known.  At
long wavelengths $(h\nu\ll kT_d)$, the multi-temperature spectrum
$\langle B_\nu\rangle (T_d)$ reduces to the Rayleigh-Jeans part of a
single blackbody with the same $T_d$.  At short wavelengths $(h\nu \gg
kT_d)$, the mean spectrum exceeds the Wien tail of a single blackbody
at the mean temperature.

\vskip-9pt
\subsubsection{2.3.5. Inputs from Galaxy Surveys}
\vskip-3pt

The formulae developed above allow us to input the observed emissivity
$E_{\nu o}$ from optical imaging and redshift surveys into our models
of cosmic chemical evolution.  The observable $E_{\nu o}$ specifies
the comoving rate of star formation $\dot\Omega_s$ through the
relation
$$
\dot\Omega_s(z)={1\over1-A_\nu(z)}\biggl[{E_{\nu o}(z)\over C_\nu(z)}
\biggr], \eqno(43)
$$
where $C_\nu$ converts between the intrinsic stellar emissivity and
star formation rate and $A_\nu$ corrects for the stellar photons
absorbed by dust in the interstellar medium (eq.~24).  Because $A_\nu$
is coupled to $\Omega_d$ (eqs.~34 and 37) and $C_\nu$ is coupled to
$\dot\Omega_s$ (eq.~25), solutions to our models of cosmic chemical
evolution must be derived iteratively (see \S~3).  In the following,
we adopt the observed ultraviolet emissivity $E_{\nu o}(z)$ from
optical surveys of galaxies and estimate the dust temperature index
$\alpha$ from observations of IRAS galaxies in the local universe.
This then leaves two unknown parameters in our models: $\tau_{cV}$
mainly affects the overall amplitude of the absorption by dust and
hence the amplitude of the far-infrared background intensity, whereas
$\epsilon$ mainly affects the evolution of the dust opacity and hence
the spectral shape of the background intensity.  Thus, $\tau_{cV}$ and
$\epsilon$ can be constrained by measurements of the cosmic
far-infrared background.

\hskip-2.8pt
Estimates of the observed ultraviolet emissivity $E_{\nu o}(z)$ are
now available over the redshift interval $0.2\lesssim z\lesssim 4.5$,
from the Canada-France Redshift Survey (CFRS, Lilly \etal 1995) and
the Hubble Deep Field (HDF, Williams \etal 1996).  Table~2 lists the
estimates of $E_{\nu o}(z)$ at several redshifts for an assumed
cosmology with $\Lambda=0$, ${q_0=1/2}$, and $H_0=50$~km~s$^{-1}$~
Mpc$^{-1}$.  The entries at $z=0.35-0.875$ were derived by Lilly
\etal (1996) from the rest-frame 2800~{\AA} luminosities of about 600
galaxies in the CFRS sample with spectroscopically confirmed redshifts
over $0<z <1$.  The entries at $z=1.25-1.75$ were derived by Connolly
\etal (1997) from the rest-frame $2800$~{\AA} luminosities of some 200
galaxies selected in the HDF images with photometrically estimated
redshifts over $1<z<2$.  The two entries at $z=2.75$ and $z=4.0$ were
derived by Madau \etal (1996, 1998): the former is based on 69
ultraviolet-dropout galaxies in the HDF images with color-estimated
redshifts of ${2.0\lesssim z\lesssim3.5}$, and the latter is based on
14 blue dropout galaxies with color-estimated redshifts of
${3.5\lesssim z\lesssim4.5}$.  The ultraviolet emissivity at $z=0$ is
not directly observed.  Instead, we use the mean $B$-band luminosity
density of local galaxies from Ellis \etal (1996), listed as the first
entry of $E_{\nu o}(z)$ in Table~2.  The values of $E_{\nu o}$ derived
by these authors all include modest extrapolations beyond the observed
range of luminosities using a Schechter function with $\alpha\approx
-1.3$.  The quoted statistical uncertainties in $E_{\nu o}$ are
$20-30$\% at $z\lesssim1$ and $40-60$\% at $z\gtrsim1$.

We determine the temperature index $\alpha$ of the dust by fits to the
observed infrared emissivity of the local universe from the IRAS
survey.  The results are shown in Figure 6, where $E_{d\nu}(0)$ is
normalized at $\lambda_1=100~\mu$m, so that the ratio of $E_{d\nu}(0)/
E_{d\nu_1}(0)$ depends mainly on $\alpha$.  The data points were
derived by Soifer \& Neugebauer (1991) from complete flux-limited
samples ($z\le0.08$ and $\langle z\rangle =0.006$) at $\lambda=12$,
25, 60, and 100~$\mu$m.  The plotted errors reflect statistical
uncertainties in the corresponding luminosity functions tabulated by
Soifer\break

\vskip19pt
\psfig{file=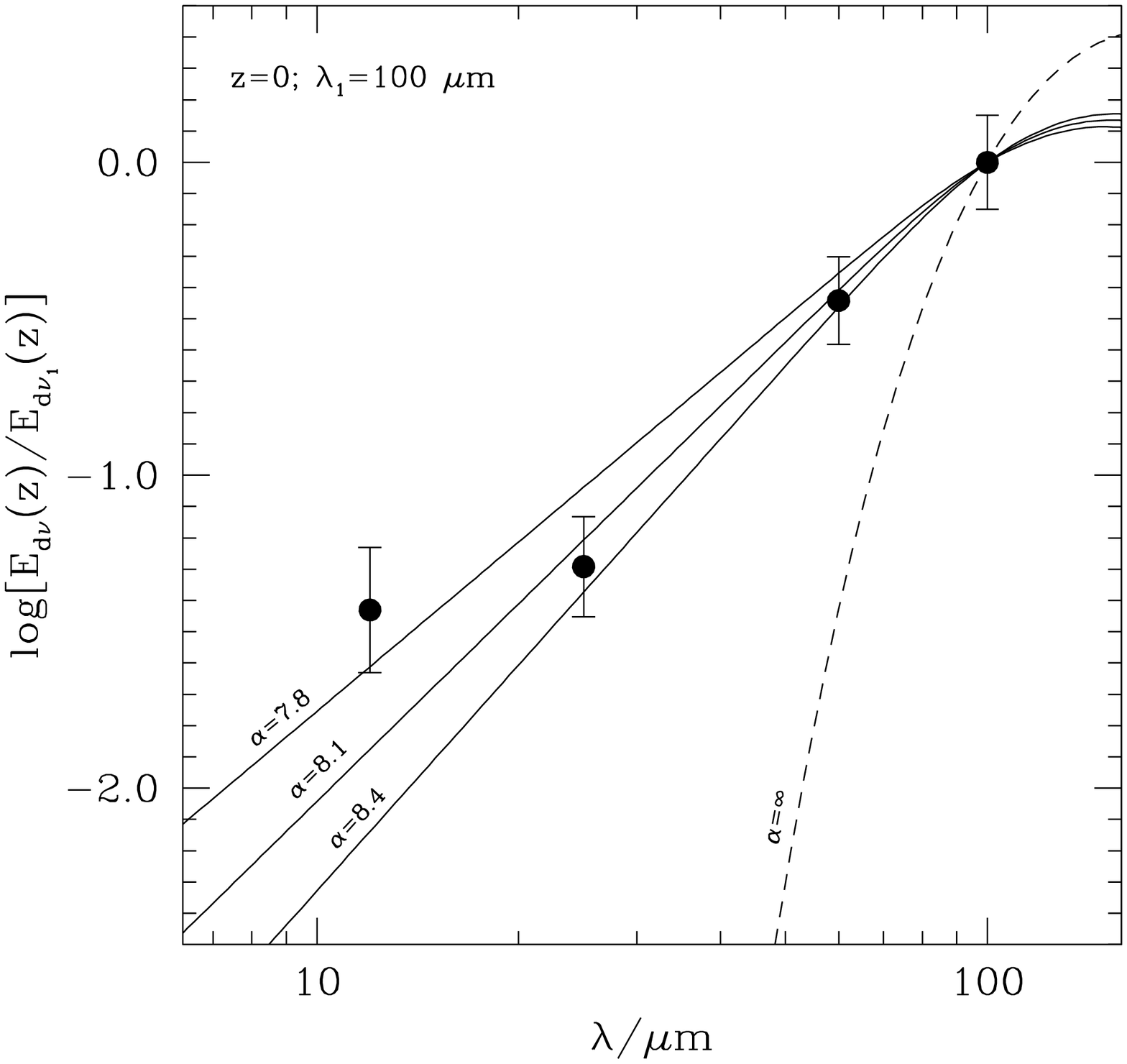,width=3.23in}
\vskip-12pt
F{\sc IG}.~6.---Normalized infrared emissivity at $z=0$.  The
data points were derived by Soifer \& Neugebauer (1991) from IRAS
observations of nearby galaxies at $\lambda=12$, 25, 60, and
100~$\mu$m.  The solid curves represent the multi-temperature models
of dust emission with the indicated values of the temperature index
$\alpha$, while the dashed curve represents a single blackbody.

\noindent
\& Neugebauer (1991).  The solid curves show the
multi-temperature models with the three indicated values of $\alpha$,
while the dashed curve shows a single blackbody (with
$\alpha=\infty$).  All curves are computed with $\kappa_a\propto
\nu^2$, an approximation at $\lambda\gtrsim10~\mu$m for LMC-type dust
(\S~2.2.5), and $T_d(0)= 16.5~$K, a typical mean temperature of dust
at $z=0$ obtained in our models (\S~3.3), but these curves are
insensitive to the exact value of $T_d$, varying by roughly 30\% at
$12~\mu$m for $T_d\approx10-25~$K.  In the context of this work,
namely, using the cosmic infrared background to correct for any
missing starlight in the optical data, it is important to have a good
account of the total infrared energy of the dust emission.  A single
blackbody, if adopted, would underestimate the mid-infrared emission,
missing roughly 50\% of the total infrared energy.  We adopt the
multi-temperature spectrum with $\alpha=8.1$ in all of our
calculations.  This still underestimates the emission near $12~\mu$m.
The observed excess is believed to be contributed by very small grains
(including polycyclic aromatic hydrocarbons), which can be excited by
single photon hits and produce non-thermal emission with complex
emission features (D\'esert \etal 1990).  Nevertheless, the
mid-infrared excess near 12~$\mu$m contributes less than 10\% of the
total infrared energy and hence should have little effect on our fits
to the cosmic infrared background.

\section {3. RESULTS}

Our models of cosmic chemical evolution are specified by equations
(4)-(6).  These equations link the history of metal enrichment to the
histories of star formation and gas consumption in galaxies and the
baryon flow between the interstellar and intergalactic media.  We
adopt two observational inputs to solve the models: (1) the observed
comoving density of HI, $\Omega_{{\rm HI}o}$, from damped Ly$\alpha$
surveys, and (2) the observed ultraviolet emissivity of galaxies,
$E_{\nu o}$, from optical imaging and redshift surveys.  The first is
related to the comoving density of gas, $\Omega_g$, (eq.~19),
including a correction factor, $Q$, for the absorbers missing from
existing damped Ly$\alpha$ surveys that is self-consistently coupled
to the comoving density of dust (eq.~17).  The second is related to
the comoving rate of star formation, $\dot\Omega_s$, (eq.~43), which
depends on a conversion factor, $C_\nu$, between the intrinsic
emissivity and star formation rate (with the aid of stellar population
synthesis models in eq.~25) and a correction factor, $A_\nu$, for
stellar radiation absorbed by dust within galaxies that is again
self-consistently coupled to the comoving density of dust (eq.~37).
The background intensity, $J_\nu$, at the present epoch is given by
equation (21), including the radiation from both stars (eq.~24) and
dust (eq.~39).

\footnote{}{\vskip5.5pt\vskip-12pt
 \tabbot{1}{M{\sc ODEL} I{\sc NPUTS}}
 {
 #\hfil&#\hfil&#\hfil\cr
 \tabhead{
  Parameter&Value&Source\cr}
 \tabbody{
  Present mean metallicity      & $Z(0)=Z_\odot=0.02$  &solar value \cr
  Mean dust-to-metals ratio     & $d_m=0.45\pm0.1$     &fit - Fig.~1\cr
  Slope of bright quasar counts & $\beta=2.0\pm0.1$    &fit - Fig.~2\cr
  Index of gamma distribution   & $\gamma=1.2\pm0.1$   &fit - Fig.~3\cr
  Norm  of gamma distribution   & $f_*=0.03\pm0.01$    &fit - Fig.~3\cr
  Index of temperature distribution &$\alpha=8.1\pm0.3$&fit - Fig.~6\cr
  Opacities of grains $\kappa_e$ \& $\kappa_a$
                                     & LMC-type dust &model - Fig.~4\cr
  Population synthesis $F_\nu$ \& $R$
                   & Salpeter IMF, $0.1-100~M_\odot$ &model - Fig.~5\cr
  Observed density of HI $\Omega_{{\rm HI}o}(z)$
                       & Data - absorption surveys &listed in Tab.~2\cr
  Observed UV emissivity $E_{\nu o}(z)$
                          & Data - optical surveys &listed in Tab.~2\cr
          &\hskip56pt(1.13,1.0)&minimum $J_\nu$ fit\cr
  Small-scale clumpy ISM model & $(\tau_{cV},\epsilon)=\biggl\{$
          (1.06,0.4)&best    $J_\nu$ fit\cr
          &\hskip56pt(1.13,0.1)&maximum $J_\nu$ fit\cr}
 }
\tabnote{NOTE.---The present mean metallicity $Z(0)$ is taken to be the
solar value (\S~2.1).  The dust-to-metals mass ratio $d_m$ is the mean
value in nearby galaxies and damped Ly$\alpha$ systems at $0.7\le z\le
2.8$ (\S~2.1).  The bright-end slope $\beta$ of the luminosity function
of quasars is fitted from the observed counts at $B\le20$ (\S~2.2.3).
The gamma distribution parameters $\gamma$ and $f_*$ for the
distribution of HI column densities are fitted from surveys of damped
Ly$\alpha$ systems (\S~2.2.3).  The dust temperature index $\alpha$ is
estimated from the mean infrared colors of local galaxies (\S~2.3.5).
The extinction and absorption opacities of dust grains, $\kappa_e$ and
$\kappa_a$, are derived from the graphite-silicates model that fits
the observed LMC extinction curve (\S~2.2.5).  The population
synthesis spectrum $F_\nu$ and returned fraction $R$ are computed from
the latest version of the Bruzual-Charlot models with a Salpeter IMF
truncated at 0.1 and 100~$M_\odot$ (\S~2.3.2).  The present visual
optical depth $\tau_{cV}$ of small-scale dust clumps and the evolution
index $\epsilon$ of the clumps relative to the global chemical
enrichment history of galaxies are determined from fits to the DIRBE
and FIRAS measurements of the cosmic infrared background (\S~3.4).}  }

Table~1 lists all the input parameters of our models. The only
adjustable parameters are the present visual optical depth $\tau_{cV}$
of small-scale clumps of dust and the index $\epsilon$ that specifies
the evolution of these clumps relative to the mean comoving density of
dust (\S~2.3.3).  Since the obscuration, absorption, and emission by
dust are interrelated, all the equations in our models are coupled.
Iterative solutions are obtained as follows.  Given $\tau_{cV}$ and
$\epsilon$, the initial set of $\Omega_g$ and $\dot\Omega_s$ (and
hence all other quantities as functions of redshift in our models) is
computed with $Q= 1$, $A_\nu=0$, and $C_\nu$ from a constant
$\dot\Omega_s$.  The values of $Q$, $A_\nu$, and $C_\nu$ are then
updated from this initial solution so that the next solution can be
computed.  This iterative process is repeated until $\Omega_g$ and
$\dot\Omega_s$ (and hence all the other quantities) converge at all
redshifts.  The stellar and dust emissivities, $E_{s\nu}$ and
$E_{d\nu}$, as functions of $z$, and the background intensity,
$J_\nu$, at $z=0$ are then computed from the converged solutions.
Fits of the far-infrared part of $J_\nu$ to the background
measurements then give the values of $\tau_{cV}$ and $\epsilon$.
These parameters are not strongly correlated; $\tau_{cV}$ determines
mostly the amplitude of the far-infrared background, while $\epsilon$
controls mostly the spectral shape of the background.  In addition to
the best-fit solution, we also present the 95\% confidence minimum and
maximum solutions to bracket the observational uncertainties in the
infrared background measurements (described in detail in \S~3.4).  The
corresponding values of $\tau_{cV}$ and $\epsilon$ are listed in
Table~1.

\subsection {3.1. Interstellar Gas and Star Formation}

The comoving density of interstellar gas $\Omega_g$ and the comoving
rate of star formation $\dot\Omega_s$ are the two key quantities from
which all other quantities in our models can be calculated.  Table~2
lists selected solutions ($\Omega_g$, $Z$, $\dot\Omega_s$, $A_\nu$),
together with the input data $\Omega_{{\rm HI}o}$ and $E_{\nu o}$.  We
have adopted discrete input data, because $\Omega_{{\rm HI}o}(z)$ at
different $z$ is obtained from samples of quasars selected at
different effective wavelengths $\lambda_e$, while $E_{\nu o}(z)$ at
different $z$ is measured at different rest-frame wavelengths
$\lambda$.  Accordingly, $Q(z)$ is evaluated at the same redshifts as
$\Omega_{{\rm HI}o}(z)$, while $A_\nu (z)$ and $C_\nu(z)$ (averaged
through a $\pm10$\% wavelength box filter) are evaluated at the same
redshifts as $E_{\nu o}(z)$.  The discrete values of $Q(z)$,
$A_\nu(z)$, and $C_\nu(z)$, along with $\Omega_{{\rm HI}o} (z)$ and
$E_{\nu o}(z)$, are then sufficient to define the pointwise solutions
for $\Omega_g(z)$ and $\dot\Omega_s(z)$.  To derive other functions of
redshift, we interpolate in $\Omega_g(z)$ and $\dot\Omega_s(z)$ using
the form $\log\Omega(z)\propto z$.  With other interpolation schemes,
such as $\log\Omega(z)\propto\log(1+z)$, the results would differ
typically by a few percent, reaching only 10\% at $z\approx4$.

\footnote{}{\vskip-12pt\def\s{\hskip6pt}
 \tabbot{2}{S{\sc ELECTED} S{\sc OLUTIONS}}
 {
 \hfil#\hfil&\hfil#\hfil&\hfil#\hfil&\hfil#\hfil&\hfil#\hfil\cr
 \tabhead{
  $z$&$\Omega_{{\rm HI}o}(z)$&$\lambda_e$&$\Omega_g(z)$&$Z(z)/Z_\odot$\cr}
 \tabbody{
 0.00&(0.25)&5500&$0.50\pm0.12$&(1.000)\cr
 0.64&$0.54\pm0.31$&5500&2.28\s(1.90,\s2.45)&0.511\s(0.473,\s0.524)\cr
 1.89&$1.58\pm0.92$&5500&6.58\s(4.04,\s8.43)&0.102\s(0.076,\s0.111)\cr
 2.40&$2.15\pm0.70$&5500&6.88\s(4.23,\s9.68)&0.060\s(0.036,\s0.070)\cr
 3.17&$2.34\pm1.14$&5500&4.66\s(3.65,\s6.65)&0.026\s(0.013,\s0.040)\cr
 4.01&$1.44\pm0.58$&6500&2.15\s(1.99,\s2.63)&0.016\s(0.008,\s0.035)\cr
 \noalign{\vskip6pt\hrule\vskip6pt}
  $z$&$\log E_{\nu o}(z)$&$\lambda$&$\dot\Omega_s(z)$&$A_\nu(z)$ \cr
 \noalign{\vskip6pt\hrule\vskip6pt}
 0.000&$19.47\pm0.10$&4400&0.60\s(0.66,\s0.63)&0.468\s(0.486,\s0.487)\cr
 0.350&$18.89\pm0.07$&2800&1.97\s(2.63,\s1.86)&0.645\s(0.718,\s0.627)\cr
 0.625&$19.21\pm0.08$&2800&4.88\s(6.52,\s4.23)&0.688\s(0.758,\s0.650)\cr
 0.875&$19.53\pm0.15$&2800&11.2\s(15.0,\s9.66)&0.691\s(0.760,\s0.651)\cr
 1.250&$19.69\pm0.15$&2800&17.0\s(19.9,\s15.3)&0.671\s(0.704,\s0.642)\cr
 1.750&$19.59\pm0.15$&2800&11.1\s(9.13,\s10.6)&0.630\s(0.507,\s0.635)\cr
 2.750&$19.42\pm0.15$&1500&14.7\s(4.72,\s20.5)&0.823\s(0.472,\s0.874)\cr
 4.000&$19.02\pm0.20$&1500&2.97\s(1.21,\s7.29)&0.637\s(0.118,\s0.848)\cr
}}
\tabnote{NOTE.---Input quantities are listed in the form observation
$\pm$ $1\sigma$ uncertainty.  Output quantities are listed in the form
best (minimum, maximum) solutions allowed by the cosmic far-infrared
background.  All entries pertain to $\Lambda=0$, $q_0=1/2$, and $H_0=
50$~km~s$^{-1}$~Mpc$^{-1}$; $\Omega_{{\rm HI}o}$ and $\Omega_g$ in
units of $10^{-3}$, $\lambda_e$ and $\lambda$ in units of \AA, $E_{\nu
o}$ in units of W~Hz$^{-1}$~Mpc$^{-3}$, $\dot\Omega_s$ in units of
$10^{-4}$~Gyr$^{-1}$.  The relation between the total and net star
formation rates is $\dot\Omega_*(z)=(1-R)^{-1}\dot\Omega_s(z)$, with
$R= 0.33$, 0.32, 0.31, 0.30, and 0.28 at $z=0$, 1, 2, 3, and 4,
respectively.} }

Figure 7 shows the comoving density of interstellar gas and the
comoving rate of star formation as functions of redshift.  The filled
circles are the model solutions for $\Omega_g(z)$ and $\dot\Omega_s
(z)$, connected by the solid, short-dash, and long-dash curves to
distinguish the best, minimum, and maximum (95\% confidence)
solutions, respectively.  In the upper panel, the open circles with
error bars are the observed comoving density of neutral gas in damped
Ly$\alpha$ systems, $\mu\Omega_{{\rm HI}o}$.  The obscuration
correction to $\Omega_g$ (from the open to filled circles) is $Q=1.5$,
$2.5-3.7$, and $1.3-2.5$ at $z=0$, 1, and 3, where the ranges include
the three solutions.  This correction is large at $z\approx1$, when
the comoving density of dust is near its highest value.  The peak in
$\Omega_g$ occurs at $z\approx2.4$, just before gas is consumed faster
by star formation than it is replenished by inflow.  In the lower
panel, the open circles with error bars are the observed (i.e.,
optically inferred) star formation rates derived in the absence of
dust absorption, $E_{\nu o}/C_{\nu}$.  The absorption correction to
$\dot\Omega_s$ (from the open to filled circles) is $(1-A_\nu)^{-1}=
2.4-2.5$, $2.8-3.9$, and $1.7-7.8$ at $z=0$, 1, and 3.  All three
solutions for $\dot\Omega_s(z)$ show a rapid decrease with respect to
time at $z\lesssim1$ and an increase with time at $z\gtrsim3$.  The
best and maximum solutions remain roughly flat over $1\lesssim z
\lesssim 3$, while the minimum solution still shows an increase with
time over this redshift range.

The star formation rates in our models differ among the three
solutions by less than 30\% at $z\lesssim2$ but by a factor of three at
$z\gtrsim3$.  The similarity at low redshifts is a consequence of the
constraint imposed by the far-infrared background.  In our models, the
mean comoving density of dust is small at high redshifts, and hence
substantial absorption occurs only if the dust is extremely clumpy, as
would be the case for heavily embedded stars.  Indeed, we find that
the minimum solution corresponds to an instantaneous mixing of dust in
the interstellar medium (i.e., small-scale clumps follow the evolution
of the whole interstellar medium), whereas the maximum solution
corresponds roughly to an incomplete mixing of dust (i.e., small-scale
clumps are nearly independent of the evolution of the rest of the
interstellar medium), with the best solution representing an
intermediate case.  Therefore, the different solutions derived here
reflect different evolution of the small-scale inhomogeneity of dust
in galaxies.

{\topinsert\null\vskip-12pt
  \psfig{file=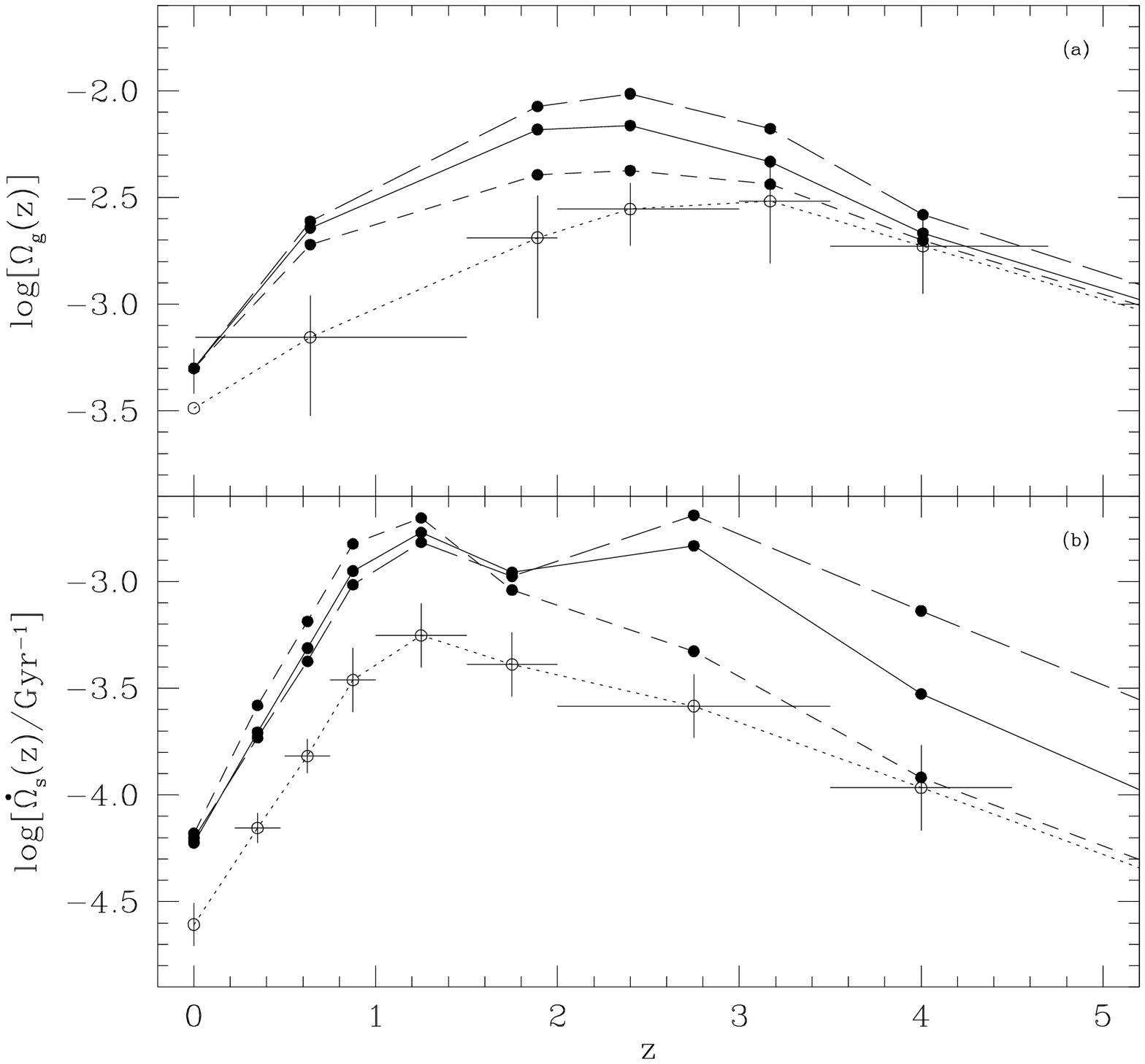,width=\h@size,silent=yes}
  \vskip12pt\edef\h@old{\the\hsize}\hsize=\h@size
  \vskip-48pt
  F{\sc IG}.~{7}.---{Comoving density of interstellar gas (upper
panel) and comoving rate of star formation (lower panel).  The filled
circles are the pointwise solutions, connected by the solid,
short-dash, and long-dash curves to distinguish the best, minimum, and
maximum solutions, respectively.  In the upper panel, the open circles
with error bars, connected by the dotted curve, are the observed
comoving density of neutral gas in damped Ly$\alpha$ systems.  In the
lower panel, the open circles with error bars, connected by the dotted
curve, are the observed ultraviolet emissivities when expressed as
star formation rates (without correction for absorption by dust).}
  \vskip24pt\vskip-5.23433pt
  \hsize=\h@old\endinsert\vfuzz=0.5pt}

\subsection {3.2. Chemical Enrichment of the Interstellar Medium}

Figure 8 shows the mean abundance, $Z$, and comoving density,
$\Omega_m$, of heavy elements in the interstellar medium as functions
of redshift in our models.  Again, the solid, short-dash, and
long-dash curves represent the best, minimum, and maximum solutions,
respectively.  The data points at $z\ge1.5$ were derived by Pettini
\etal (1997b) from the observed abundances of Zn in damped Ly$\alpha$
systems, while the data point at $z=0.77$ was derived by Boiss\'e
\etal (1998) by combining their own observations with those of Pettini
\etal (1997b).  These are the HI column density weighted mean
metallicity and hence are the correct measure of metallicity to be
compared with our models.  The uncertainty in the data is still large
and no attempt is made here to correct for a preferential selection of
the low-metallicity absorbers that would bias the observed mean.  We
emphasize, however, that this bias could potentially be large at
$z\approx1$, where the comoving density of dust is high.
Nevertheless, all three solutions for the mean metallicity in our
models are consistent within the uncertainties with the observations,
especially if the data point at $z=0.77$ is regarded as a lower limit.
The present mean metallicity $Z(0)$, taken here to be the solar value,
implies an effective yield of $y=0.45Z_\odot$, nearly the same in the
three solutions.  The peak in $\Omega_m$ at $z\approx0.8$ is due to a
combination of the rise in $Z$ and the decline in $\Omega_g$ toward
$z=0$.  The decrease in $\Omega_m$ toward $z=0$ is caused partly by
the fact that star formation starts to consume more metals than it
contributes to the interstellar medium (at $z\lesssim 0.8$, when $Z$
exceeds $y$), and partly by some outflow of the interstellar metals to
the intergalactic medium (see \S~4.4).

\vskip28pt
\psfig{file=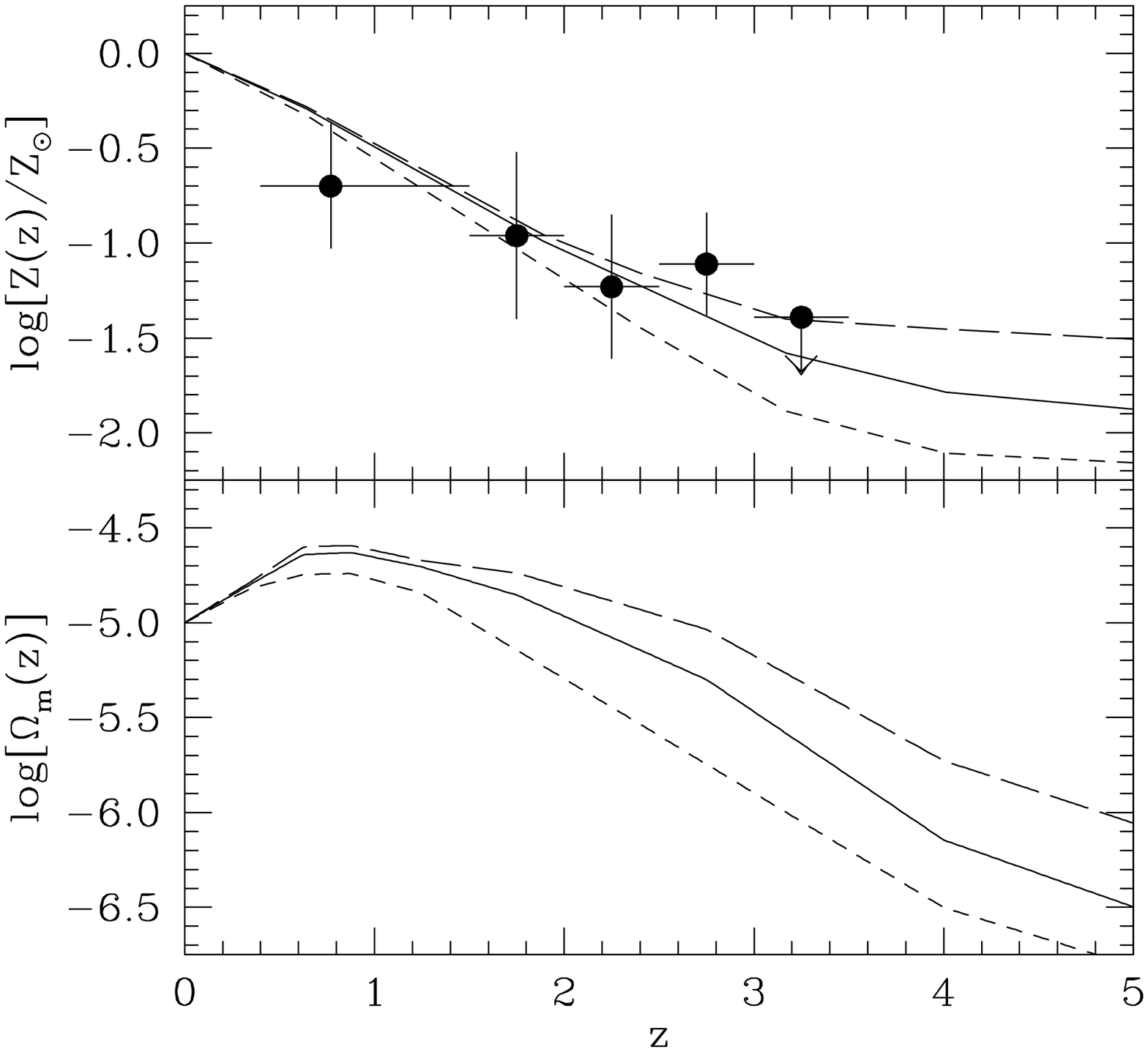,width=3.23in}
\vskip-12pt
F{\sc IG}.~8.---Mean metallicity of interstellar gas in galaxies
in units of the solar value (upper panel) and comoving density of
heavy elements in galaxies (lower panel).  The solid, short-dash, and
long-dash curves are the best, minimum, and maximum solutions in our
models.  In the upper panel, the data points at $z\ge1.5$ are from
Pettini \etal (1997b), while the data point at $z=0.77$ is from
Boiss\'e \etal (1998).

The histories of metal enrichment in the three solutions presented
here differ mainly at high redshifts and are caused by the different
histories of star formation.  This indicates that the observed mean
metallicity in the high-redshift damped Ly$\alpha$ systems should be a
useful constraint on star formation at high redshifts.  In particular,
if the global rates of star formation were much higher at $z\gtrsim3$
than in the maximum solution presented here, the interstellar
metallicity would be larger than that observed in the damped
Ly$\alpha$ systems, unless a significant outflow had occurred before
$z\approx3$.

\subsection {3.3. Evolution of Stellar and Dust Emissivities}

Figure 9 shows the emissivity $E_{\nu}(z)$ times frequency $\nu$ as a
function of redshift at several rest-frame wavelengths.  The data are
taken from Lilly \etal (1996) for 0.28, 0.44, and 1.0~$\mu$m at
$z=0.35 $, 0.625, and 0.875, Madau \etal (1998) for 0.15~$\mu$m at
$z=2.75$ and 4.0, Connolly \etal (1997) for 0.28 and 0.44~$\mu$m at
$z=1.25$ and 1.75, Ellis \etal (1996) for 0.44~$\mu$m at $z=0$,
Gardner \etal (1997) for 2.2~$\mu$m at $z=0$, and Soifer \& Neugebauer
(1991) for 25, 60, and 100~$\mu$m at $z=0$.  The curves are the three
solutions for the cosmic emissivity in our models.  The results at
$\lambda=0.15-2.2~\mu$m represent the direct but attenuated starlight,
while those at $\lambda=25-100~\mu$m represent the starlight
reradiated by dust.  The data that were input to the models are
plotted as open circles, while those that are output are plotted as
filled circles.  Evidently, our models reproduce remarkably well even
data that were not used as input.  In particular, the agreement at
$\lambda=1.0-2.2~\mu$m, where the absorption by dust is relatively
small, is an indication that the Salpeter IMF is reasonable.  A
steeper IMF, such as a Scalo IMF, would produce too much 2.2~$\mu$m
light at $z=0$ (by a factor of two).  The agreement with the observed
$25-100~\mu$m emissivity at $z=0$ indicates that the evolution of the
small-scale clumpiness of dust in the models is reasonable.  A
substantially different evolution from that presented here might give
an acceptable fit to the integrated infrared background but would not
necessarily match the local infrared emissivity at the same time.

{\topinsert\null\vskip-12pt
  \psfig{file=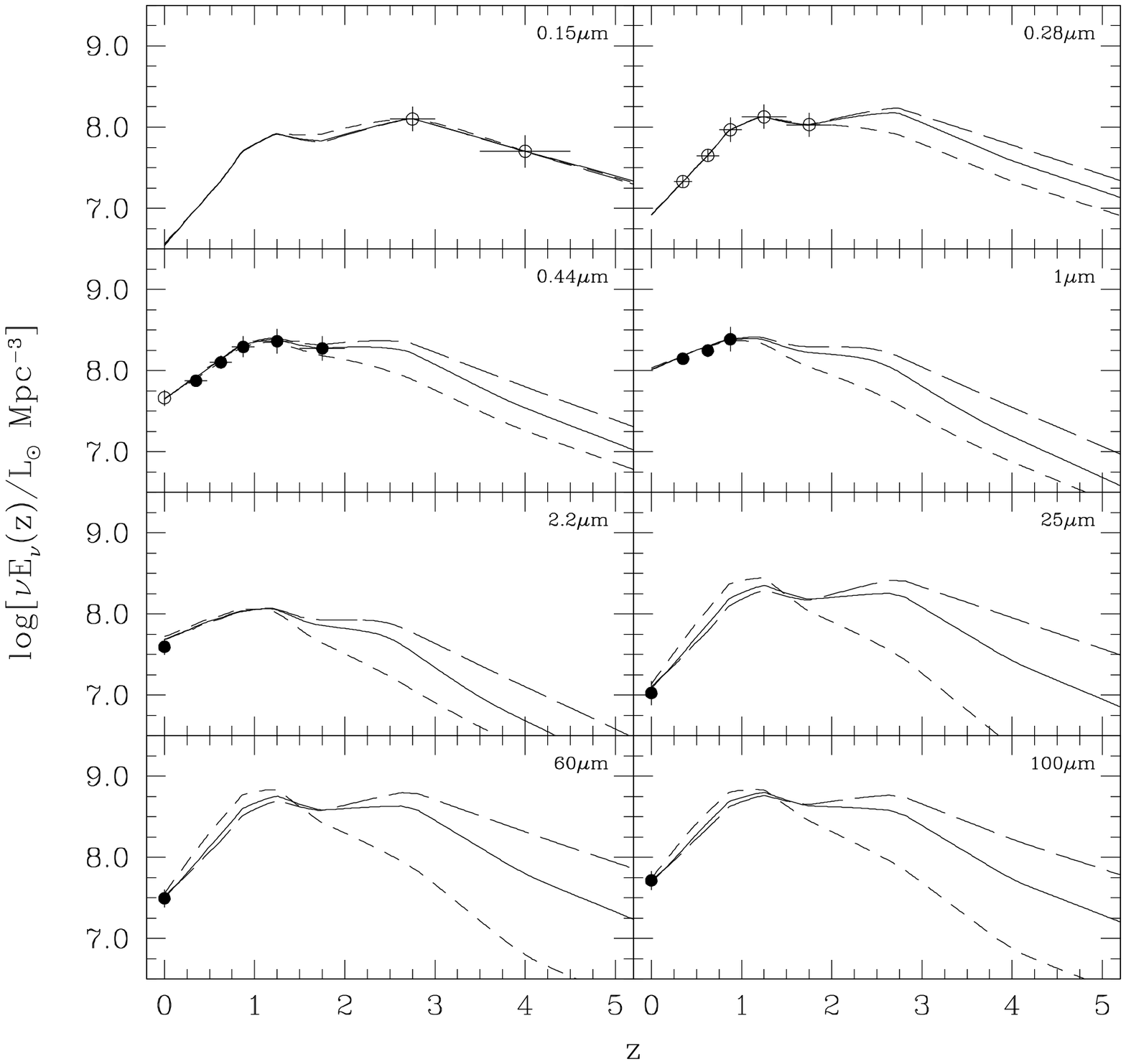,width=\h@size,silent=yes}
  \vskip12pt\edef\h@old{\the\hsize}\hsize=\h@size
  \vskip-48pt
  F{\sc IG}.~{9}.---{Comoving emissivity $E_\nu(z)$ times frequency
$\nu$ as a function of redshift.  The solid, short-dash, and long-dash
curves are the best, minimum, and maximum solutions in our models.
The data points are from Madau \etal (1998) for 0.15~$\mu$m; Lilly
\etal (1996) for 0.28, 0.44, and 1.0~$\mu$m at $z=0.25-1.0$; Connolly
\etal (1997) for 0.28 and 0.44~$\mu$m at $z=1-2$; Ellis \etal (1996)
for 0.44~$\mu$m; Gardner \etal (1997) for 2.2~$\mu$m; and Soifer \&
Neugebauer (1991) for 25, 60, and 100~$\mu$m.  All wavelengths are in
the rest frame at the indicated redshifts.}
  \vskip24pt\vskip-5.23433pt
  \hsize=\h@old\endinsert\vfuzz=0.5pt}

The mean temperature of dust, plotted in the upper\break

\vskip17pt
\psfig{file=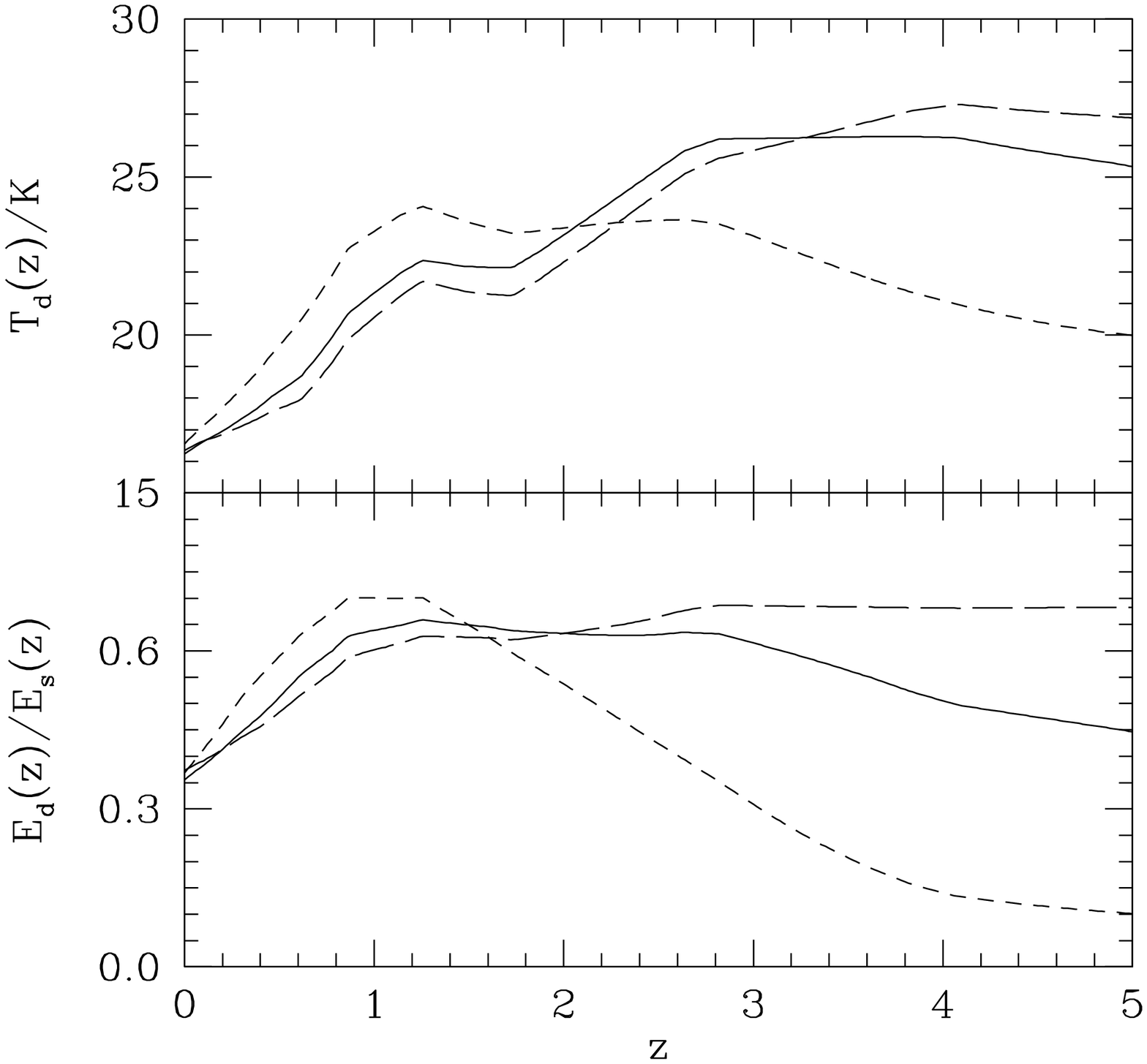,width=3.23in}
\vskip-12pt
F{\sc IG}.~10.---Mean temperature of interstellar dust (upper
panel) and ratio of bolometric dust emissivity to bolometric stellar
emissivity (lower panel).  The solid, short-dash, and long-dash curves
represent the best, minimum, and maximum solutions, respectively.

\noindent
panel of Figure~10,
varies from $T_d=21-27~$K at $z=4$ to $T_d=22-24~$K at $z=2$ and then
decreases to $T_d=16.5~$K at $z=0$.  In our models, dust grains heated
to different temperatures (at least above a finite temperature
$T_c\approx0.9T_d$) contribute different amounts to the bolometric
dust emission: 50\% by dust with $T\approx(0.9 -1.6)T_d$, 25\% by dust
with $T\approx(1.6-3.0)T_d$, and 25\% by dust with $T\gtrsim3.0T_d$,
with little dependence on redshift.  The ratio of the bolometric dust
emission to the bolometric stellar emission, defined by $E_d/E_s=\int
d\nu E_{d\nu}/\int d\nu E_{s\nu}$ and plotted in the lower panel of
Figure~10, varies from $E_d/E_s=14-68$\% at $z=4$ to $E_d/E_s=54-64$\%
at $z=2$ and then decreases to $E_d/E_s=36$\% at $z=0$.  Evidently,
the dust absorbs and reradiates a significant fraction of the stellar
energy.  Moreover, since $\tau_c$ is typically larger than $\tau_t$,
most of this absorption occurs in small-scale clumps, rather than the
large-scale distribution of dust in the interstellar medium.  This is
consistent with the general notion that the young stars responsible
for most of the energy production are typically located in dusty
star-forming regions.

\subsection {3.4. Cosmic Background Radiation}

Figure 11 shows the background intensity $J_\nu$ at $z=0$ times
frequency $\nu$ as a function of wavelength $\lambda$.  The circles at
140 and 240~$\mu$m are the DIRBE detections; the errors include both
statistical and systematic uncertainties in the measurements and the
removal of foreground emissions (Hauser \etal 1998).  The zigzag curve
at $150-2000~\mu$m is the weighted mean of the FIRAS detections, based
on three independent methods to remove the foregrounds (Fixsen \etal
1998).  The statistical uncertainties in the FIRAS results are 40\%,
27\%, 5\%, 6\%, 10\%, and 15\% at $\lambda=160$, 180, 240, 400, 700,
and 1000~$\mu$m, respectively (Fixsen 1998, private communication).
The solid curve is the weighted minimum-$\chi^2$ fit of the background
intensity in our models to the far-infrared/submillimeter DIRBE and
FIRAS measurements to determine $\tau_{cV}$ and $\epsilon$.  We regard
this as our ``best'' solution.  The systematic uncertainties in the
FIRAS results are not known precisely.  For this reason, two other
acceptable fits are presented as the short-dash and long-dash
curves.  These ``minimum'' and ``maximum'' solutions would be the 95\%
confidence lower and upper limits if the systematic uncertainties in
the FIRAS data were taken to be the difference between the three
independent methods in the removal of the foregrounds by Fixsen \etal
(1998).

Also shown in Figure 11 is a comparison of the backgrounds in our
models with various measurements and limits at other wavelengths.  The
squares with errors are the integrated light from observed sources at
2000~{\AA} (Armand, Milliard, Deharveng 1994) and $3600-22000 $~{\AA}
(Pozzetti \etal 1998).  The large error at 2000~{\AA} is due to the
extrapolation of the counts beyond a limiting magnitude of 18.5.  At
optical wavelengths, the slope of the counts suggests that very faint
uncounted galaxies contribute little to the background.  Thus, we
regard the integrated light from the observed sources as tantamount to
a measurement of rather than a lower limit on the background.  The
open circle is a tentative detection of the extragalactic background
at $\lambda=3.5~\mu$m from the DIRBE (2.2 and 3.5~$\mu$m) maps and an
assumed value of the background at $\lambda=2.2~\mu$m by Dwek \&
Arendt (1998).  Various (downward and upward) arrows are observational
(upper and lower) limits: the DIRBE $2\sigma$ upper limits at
$1.25-100~\mu$m (Hauser \etal 1998), the high galactic latitude upper
limits at 1000~{\AA} (Holberg 1986) and 1700~{\AA} (Bowyer 1991), the
sky surface photometry upper limits at 4400~{\AA} (Toller 1983) and
5100~{\AA} (Dube, Wickes, \& Wilkinson 1979), the lower limits in the
no-evolution case from IRAS galaxy counts at 25, 60, and 100~$\mu$m
(Hacking \& Soifer 1991), the upper limits at $10-100~\mu$m from the
fluctuation analysis of the DIRBE maps (Kashlinsky, Mather, \&
Odenwald 1996), and the upper limits at $5-25~\mu$m (Stanev \&
Franceschini 1998) and at $1-50~\mu$m (stepped line, Biller \etal
1998) from the lack of attenuation in the TeV gamma-ray spectrum of
Mrk 501.  Evidently, our models agree well with a wide variety of
measurements of and limits on the background intensity over four
decades in wavelengths.

{\topinsert\null\vskip-12pt
  \psfig{file=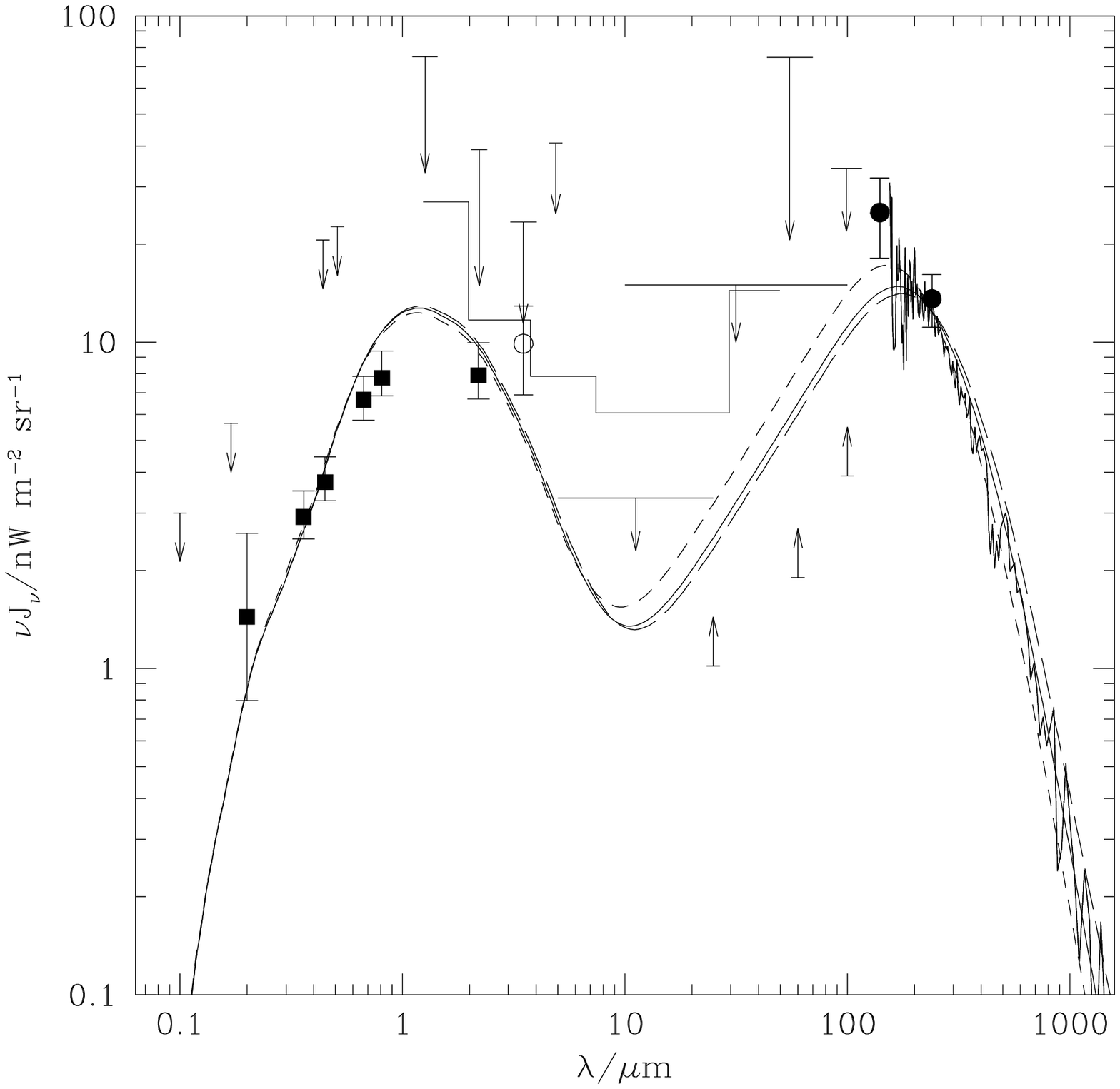,width=\h@size,silent=yes}
  \vskip12pt\edef\h@old{\the\hsize}\hsize=\h@size
  \vskip-48pt
  F{\sc IG}.~{11}.---{Extragalactic background intensity $J_\nu$ times
frequency $\nu$ as a function of wavelength $\lambda$.  The circles
are the DIRBE detections of the cosmic infrared background at 140 and
240~$\mu$m (Hauser \etal 1998).  The zigzag curve is the FIRAS
detections at $150-2000~\mu$m (Fixsen \etal 1998).  The solid,
short-dash, and long-dash curves represent the best, minimum, and
maximum fits of our models to these observations.  The filled squares
are the integrated background light from galaxy counts at 2000~{\AA}
(Armand \etal 1994) and $3600-22000~${\AA} (Pozzetti \etal 1998).  The
open circle is a tentative detection of the extragalactic background
at $3.5~\mu$m (Dwek \& Arendt 1998).  The arrows and the stepped line
indicate various observational limits on the extragalactic background
radiation (see \S~3.4 for references).}
  \vskip24pt\vskip-5.23433pt
  \hsize=\h@old\endinsert\vfuzz=0.5pt}

The total energy density radiated by stars of all ages and now
contained in the extragalactic background (over $912~{\rm \AA}
\le\lambda\le5~$mm) is, in units of the present critical energy
density,
$$
\Omega_R\equiv{32\pi^2 G\over3H_0^2c^3}
\int_0^\infty d\nu J_\nu=(5.1-5.5)\times10^{-6}. \eqno(44)
$$
Of this, $54-59$\% is reprocessed by dust into the cosmic infrared
background.  For comparison, the energy density in the cosmic
microwave background radiation left by the Big Bang is about 20 times
larger than that given by equation (44) (Fixsen \etal 1996).  In these
estimates, we have ignored the unknown contribution from active
galactic nuclei.

\section {4. DISCUSSION}

We have combined our models with a wide variety of observations to
study the average evolution of the entire population of galaxies.
Before discussing some implications of our solutions (\S~4.4 and
\S~4.5), we examine some major uncertainties that could potentially
affect our results (\S~4.1 and \S~4.2) and compare them with other
estimates (\S~4.3).

All of the results presented here are based on a cosmological model
with $\Lambda=0$, $q_0=1/2 $, and $H_0=50$~km~s$^{-1}$~Mpc$^{-1}$.
These parameters may no longer be regarded as ``standard'' in light of
recent evidence, especially from high-redshift supernovae (see, for
example, Reiss \etal 1998).  However, many of the observational input
data on which our results depend were reported only for this
particular set of cosmological parameters.  A self-consistent
exploration of the effects of different cosmological parameters would
therefore require a reanalysis of the input data, particularly
$\Omega_{{\rm HI}o}$ and $E_{\nu o}$.  Such a study might prove
interesting but is beyond the scope of this paper.

\subsection {4.1. Model Uncertainties}

Our results are independent of many properties of the damped
Ly$\alpha$ systems, including their morphologies, provided only that
they represent the main reservoirs of gas for star formation.  The
existing evidence, although not yet conclusive, suggests that the
damped Ly$\alpha$ systems are galaxies of a wide variety of
morphological types (Le Brun \etal 1997).  The idea that the damped
Ly$\alpha$ systems trace the the bulk of star formation is supported
by the observation that they contain enough neutral gas at
$z\approx2-3$ to make all the stars visible today (Wolfe \etal 1995).
Such a large amount of neutral gas is not present in the local
universe, even including the contribution of low surface brightness
and dwarf galaxies (Zwaan \etal 1997).  Indeed, if the neutral gas in
damped Ly$\alpha$ systems were not converted into stars, it must have
been expelled or ionized at low redshifts.

The main approximation in our treatment of the interstellar gas in
galaxies is the neglect of molecules.  As discussed in \S~2.2.1, this
could increase $\Omega_g$ by up to a factor of two, most likely at low
redshifts.  Accordingly, $\Omega_d$ would be also be higher because it
is linked to the metallicity and the present value $Z(0)$ is fixed in
our approach.  However, as we have already emphasized, $Z(0)$ is not
known precisely, and different values of this parameter affect
$\Omega_d$ through $\Omega_g$ as well.  We have checked that these
uncertainties, comparable to the errors in the observed comoving
density of HI at low redshifts, do not strongly affect our solutions
for the global rate of star formation.  We have also checked that the
uncertainties in the input parameters ($d_m,\beta,\gamma,f_*,\alpha$)
all have minor effects on our results.

A natural consequence of dust in galaxies is that it obscures
background quasars (Ostriker \& Heisler 1984).  This effect also
causes incompleteness in samples of damped Ly$\alpha$ systems found in
the spectra of optically selected quasars (Fall \& Pei 1993).  If this
bias were ignored in our models, the resulting mean interstellar
metallicity at $z\approx2.5$ would be $3-5$ times higher than that
observed in the damped Ly$\alpha$ systems.  As discussed at length in
\S~2.2.2, our models include self-consistent corrections for missing
damped Ly$\alpha$ systems and quasars.  From our solutions for $f$ and
$f_o$ and the formulae of Fall \& Pei (1993), we estimate that the
missing fraction of optically selected quasars with $V\lesssim20$
($R\lesssim20$) is relatively small: $9-12$\% ($8-11 $\%), $12-19$\%
($10-17$\%), and $13-23$\% ($12-21$\%) at $z=2$, 3, and 4,
respectively.  An important observational constraint on obscuration
comes from counts of radio and optically selected quasars on the
assumption that they have the same intrinsic evolution.  Shaver \etal
(1996) found that the comoving densities of radio and optically
selected quasars have similar apparent evolution to $z\approx4$.
However, the samples at high redshift are very small and are easily
consistent with $40-50$\% of the optically selected quasars being
missed at $z\approx3-4$ (see Fig.~2 of Shaver \etal 1996).  The
obscuration in our models is well within this limit.

The determination of the global star formation rate from the observed
ultraviolet emissivity of galaxies involves assumptions about the IMF
and about dust (eq.~43).  In our approach, all uncertainties in the
IMF (its shape, mass cutoffs, and any redshift evolution) are combined
in the conversion factor $C_\nu(z)$ between the intrinsic emissivity
and star formation rate (\S~2.3.2).  All uncertainties related to the
dust (its opacity, evolution, and distribution) are combined in the
mean fraction $A_\nu(z)$ of photons absorbed by dust (\S~2.3.3).  Our
solutions are based on the following assumptions: (1) an IMF that is
constant in time; (2) an IMF with Salpeter form; (3) a constant
dust-to-gas ratio and a power-law relation between the optical depth
of small-scale clumpiness and mean comoving density of dust; (4)
optical opacity of grains as in LMC-type dust; and (5) infrared
opacity as in LMC-type dust.  We now discuss each assumption in turn.
The basic conclusion is that the small-scale clumpiness of the dust
and the optical opacity of grains are the major sources of uncertainty
in the determination of the global star formation history.

(1) Some of the main inputs to our models are the observed rest-frame
1500~{\AA} and 2800~{\AA} emissivities.  These models, with a constant
IMF, also reproduce as output the observed rest-frame 4400~{\AA} and
$1~\mu$m emissivities at $0.2\lesssim z\lesssim2.0$ (Fig.~9).  This
may be an indication that any evolution of the IMF is relatively weak,
at least for $z\lesssim2$.  Since absorption by dust decreases with
wavelength, the rest-frame $2.2~\mu$m emissivity of galaxies, if
observed over a similar range of redshifts, could provide a tighter
constraint on this evolution than the rest-frame 4400~{\AA} and
$1~\mu$m emissivities.

(2) The Salpeter IMF is flatter (has more massive stars) than the
Scalo IMF, the other IMF often employed in studies of stellar
populations.  The appropriateness of the Salpeter IMF over the Scalo
IMF to reproduce the observed emissivities at $z=0$ has already been
noted by Lilly \etal (1996) and Madau \etal (1998).  Repeating all of
our calculations with a Scalo IMF, we find that the solutions for
$\dot\Omega_s(z)$ are not strongly affected, except at very low
redshifts ($z\lesssim0.5$).  However, a Scalo IMF would predict too
much red light, causing a factor of roughly $2-3$ mismatch of our
models with the observed $2.2~\mu$m local emissivity and with the
$1-2.2~\mu$m extragalactic background.  We have not fine-tuned the
upper and lower mass cutoffs of the IMF.  The lower mass cutoff could
potentially affect the overall amplitude of the star formation rate,
without any strong effects on the stellar emissivity.  With a lower
mass cutoff of 0.1~$M_\odot $, the comoving density of stars at $z=0$
in our models agrees well with the empirical estimates (\S~4.3).

(3) Our models of cosmic chemical evolution represent a
self-consistent way to relate the evolution of dust to the history of
star formation on the assumption of a constant dust-to-metals ratio.
If this ratio were to vary by 50\% over $0\lesssim z\lesssim3$
(Fig.~1), it would have a minor effect on the solutions for
$\dot\Omega_s(z)$.  We have adopted a simple statistical approach to
characterize the small-scale clumpiness of dust in the interstellar
medium.  This introduces two adjustable parameters: the present value
of the optical depth of the clumps $\tau_{c}$ and an index $\epsilon$
that specifies how the clumps evolve relative to the mean comoving
density of dust.  In our models, the small-scale clumpiness of dust is
responsible for most of the absorption of starlight.  We have
presented three solutions in which the small-scale clumpiness of the
dust evolve differently.  This turns out to be a major source of
uncertainty in our solutions for the global star formation rate.

(4) The optical opacity of grains varies between galaxies by a factor
of two, even among the Milky Way, LMC, and SMC (Fig.~4).  To assess
this effect on our results, we repeated our calculations with Galactic
and SMC-type dust.  The resulting infrared background was similar
within 30\%, while the ultraviolet background at $\lambda \approx
2000~${\AA} differed by a factor of two higher (lower) for
Galactic-type (SMC-type) dust relative to the LMC-type dust.  The
emissivity $E_{\nu o}$ at rest-frame ultraviolet wavelengths increased
(decreased) by a factor of two at low redshifts and decreased
(increased) at high redshifts.  The solution for $\dot\Omega_s(z)$ was
similar at $z\lesssim2$ but a factor of two lower (higher) at
$z\gtrsim3$.  These differences are consequences of the fact that
Galactic-type (SMC-type) dust selectively absorbs less (more)
ultraviolet radiation than the LMC-type dust.  Therefore, the high-$z$
solutions of $\dot\Omega_s(z)$ are uncertain (by a factor of two) due
to the uncertainty in the optical opacity of grains, comparable to or
less than the uncertainty caused by the small-scale clumpiness of the
dust.  The solutions presented here are all based on LMC-type dust,
because this is an intermediate case between Galactic and SMC-type
dust and because, in combination with a Salpeter IMF, it seems to
account well for all the available data.

(5) The infrared opacity of grains, responsible for the dust emission,
is approximated here by a power law, $\kappa_a(\nu)\propto \nu^n$,
where the normalization and index ($n=2$) are fixed by our adopted
LMC-type dust (see Fig.~4).  The normalization has roughly the same
effect as $\tau_{cV}$ on the amplitude of $J_\nu$, while the index has
roughly the same effect as $\epsilon$ on the shape of $J_\nu$.  Since
our fits to the cosmic infrared background involve adjustment of
$\tau_{cV}$ and $\epsilon$, the uncertainties in the infrared opacity
of grains are partially absorbed by these parameters.  We have checked
that a different infrared opacity (within 50\% of the LMC-type dust)
would have only minor effects on our results (but with different
values of $\tau_{cV}$ and $\epsilon$).  However, if the normalization
were much higher (lower) than adopted here, it would require much more
(less) absorption of stellar energy to reproduce the same infrared
background, causing the optical background to be too blue (red).  A
value of $n=1.5$ would only give a marginally acceptable fit to the
far-infrared background.

\subsection {4.2. Observational Uncertainties}

We have used as input to our models the best determinations available
in 1997 of the comoving density of HI and the rest-frame ultraviolet
emissivity, $\Omega_{{\rm HI}o}$ and $E_{\nu o}$.  Any errors in
$\Omega_{{\rm HI}o}$ and $E_{\nu o}$ would propagate into the
corresponding solutions for $\Omega_g$ and $\dot\Omega_s$.  Since the
quoted statistical errors in $\Omega_{{\rm HI}o}$ and $E_{\nu o}$ are
all 60\% or less, these uncertainties do not strongly affect our
results.  In addition to the statistical errors, however, it is always
possible that the input data are plagued by unknown systematic errors
and/or biases.  After completing this work, we became aware of new
estimates of $E_{\nu o}$ at high redshifts by Steidel \etal (1999)
from a large ground-based survey of Lyman-break galaxies: $\log E_{\nu
o}=19.42\pm0.07$ at $z=3.04$ and $\log E_{\nu o}= 19.33\pm 0.10$ at
$z=4.13$, both for $\lambda=1500$~{\AA} and by integrating a Schechter
luminosity function down to $0.1~L_*$ (Steidel 1999, private
communication).  The new estimate of $E_{\nu o}$ at $z\approx4$ is
higher than the previous one by Madau \etal (1998) from the HDF by a
factor of two.  Steidel \etal (1999) suggest that the reason for this
discrepancy is that the HDF may be too small to be a fair sample of
the universe.  Another source of uncertainty in $E_{\nu o}$ is the
fact that the faint end of the luminosity function has not been well
established at high redshifts.

We have repeated all our calculations with the new estimates of
$E_{\nu o}$ by Steidel \etal (1999) at $z\ge3$ and the entries in
Table~2 at lower redshifts.  We then readjusted the parameters
$\tau_{cV}$ and $\epsilon$ in the models to match the cosmic
far-infrared background measurements.  The resulting best, minimum,
and maximum solutions for $\dot\Omega_s$ are plotted in Figure 12,
with $(\tau_{cV},\epsilon)=(0.82,0.8)$, $(0.98,1.4)$, and $(0.88,0.4)$,
respectively.  These solutions are not radically different from those
shown in Figure 7b.  The largest difference between the best solutions
in the two cases is only 50\% at $z\lesssim 4$, even though the
corresponding input values of $E_{\nu o}$ at $z\approx4$ differ by a
factor of two.  The reason for this is that our solutions for
$\dot\Omega_s$ are constrained by many observational input data and
hence the propagation of errors from $E_{\nu o}$ to $\dot\Omega_s$ is
significantly reduced (by a factor of two at $z\approx4$).  Given
these uncertainties in the input data, the global rate of star
formation is fairly well determined by our models.

\vskip22pt
\psfig{file=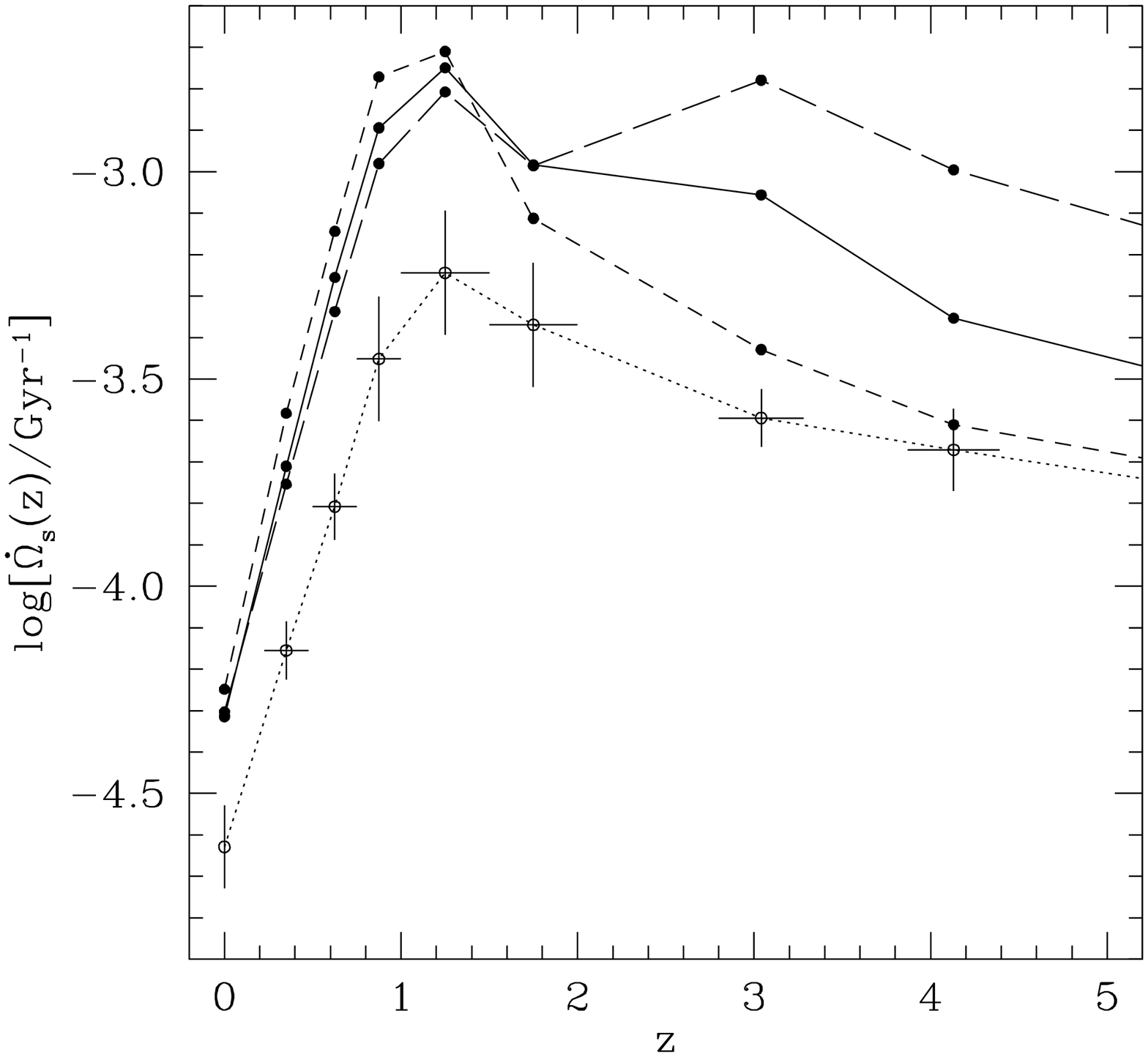,width=3.23in}
\vskip-12pt
F{\sc IG}.~12.---Comoving rate of star formation in our models
based on the modified emissivities at $z\ge3$ from Steidel \etal
(1999).  The symbols have the same meaning as in Fig.~7b.
\vskip7pt

\subsection {4.3. Comparisons with Other Results}

The global rates of star formation derived here are compared with
independent estimates made by other techniques in Figure~13.  The
solid, short-dash, and long-dash curves are respectively the best,
minimum, and maximum solutions for $\dot\Omega_*(z)$ in our models.
The dotted curves are the predictions of the inflow (upper curve) and
outflow (lower curve) models of Pei \& Fall (1995), which are based on
absorption-line observations of damped Ly$\alpha$ systems.  The
circles at $z=0$, 0.2, and 0.8, come from H$\alpha$ observations of
galaxies in a local sample (Gallego \etal 1995) and in the CFRS sample
(Tresse \& Maddox 1998; Glazebrook \etal 1999).  For consistency, we
have adopted the Glazebrook \etal (1999) conversion from the H$\alpha$
luminosity density $L_{{\rm H}\alpha}$ (corrected for dust using
H$\beta$-to-H$\alpha$ line ratios) to the star formation rate
(appropriate for the Salpeter IMF with solar metallicity), equivalent
to $(\dot\Omega_*/{\rm Gyr}^{-1})=1.0 \times10^{-43}(L_{{\rm
H}\alpha}/{\rm erg~s}^{-1}~{\rm Mpc}^{-3})$.  A similar conversion
factor is also found in our models (assuming 0.45 H$\alpha$ photons
per Lyman continuum photon for case B recombination and including the
metallicity dependence of the spectral evolution).  The squares at
$z\approx0.2-1.0$ come from ultraviolet, optical, near-infrared, and
radio observations of ISO 15~$\mu$m selected galaxies in the CFRS
survey (Flores \etal 1999).  Clearly, our solutions for the global
rates of star formation agree well with the available H$\alpha$ and
mid-infrared surveys at $z\lesssim1$.  They are also consistent with
an estimate of the star formation rate at $z\approx3$ by Hughes \etal
(1998) from submillimeter observations of the HDF with SCUBA, although
this depends on only a few detected sources and assumptions about
their dust temperatures and redshifts (see additional results from
Barger \etal 1998).  The global history of star formation derived here
is also broadly consistent with recent estimates of the redshift
distribution of SCUBA sources (Lilly \etal 1999).

In summary, we conclude that our solutions for the\break

\vskip16.5pt
\psfig{file=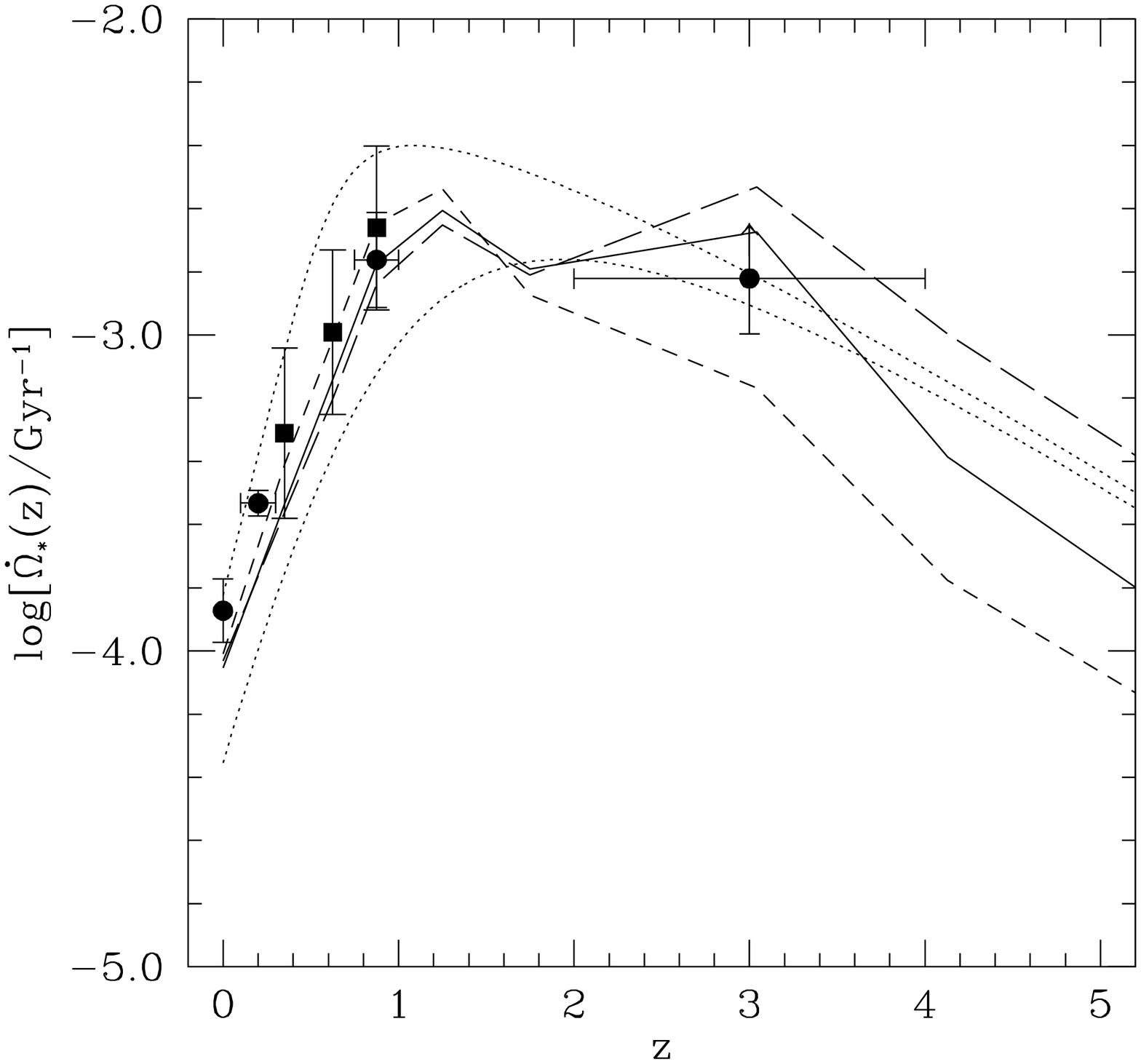,width=3.23in}
\vskip-12pt
F{\sc IG}.~13.---Comparisons of the comoving rate of star
formation in our models with estimates from recent H$\alpha$,
mid-infrared, and submillimeter observations.  The solid, short-dash,
and long-dash curves are respectively the best, minimum, and maximum
solutions derived here.  The two dotted curves are the predictions of
Pei \& Fall (1995) from models of cosmic chemical evolution.  The
circles at $z=0$, 0.2, and 0.875 are from the H$\alpha$ surveys by
Gallego \etal (1995), Tresse \& Maddox (1998), and Glazebrook \etal
(1999), respectively.  The squares are from the ISO 15$\mu$m survey by
Flores \etal (1999).  The circle with an arrow at $z\approx3$ is from
the SCUBA submillimeter observations by Hughes \etal (1998).

\noindent
global rate of star
formation are robust at $z\lesssim2$, but still uncertain by a factor
of three at $z\gtrsim3$.  The robustness at $z\lesssim2$ is due to the
tight constraint imposed by the cosmic infrared background on how much
radiation is emitted and absorbed at shorter wavelengths.  At
$z\gtrsim3$, there is little time to accumulate much radiation in the
background.  Consequently, the observed background provides only a
weak constraint on the star formation rates at high redshifts.  Our
models indicate that the rest-frame emissivity of galaxies evolves
strongly not only in the near-ultraviolet (e.g., 2800~{\AA}) but also
in the far-infrared (e.g., $100~\mu$m).  Either one of these
emissivities, if studied in isolation, would provide only a partial
tracer of star formation.  The most promising way to determine the
global rates of star formation at $z\gtrsim3$ will be to combine
future near-infrared and submillimeter surveys; the former traces the
direct radiation from stars at rest-frame near-ultraviolet
wavelengths, while the latter traces the reradiation by dust at
rest-frame far-infrared wavelengths.  For $z\gtrsim4$, stellar
emission at $\lambda=2800~${\AA} is redshifted to $\lambda\gtrsim
1.4~\mu$m, and falls within the window of the Next Generation Space
Telescope (NGST), whereas dust emission at $\lambda =100~\mu$m is
redshifted to $\lambda\gtrsim500~\mu$m, and falls within the window of
the Millimeter Array (MMA).  Thus, a combination of these two
instruments should eventually enable us to trace the global history of
star formation to high redshifts.

\vskip-0.5pt
\subsection {4.4. Evolution of Galaxies}

Our models of cosmic chemical evolution provide a global description
of the major constituents of galaxies.  Figure 14 shows the inferred
evolution of the stars, interstellar gas, baryons in galaxies, and
baryon flow between the interstellar and intergalactic media.  The
data point $\Omega_s=(4.9^{+3.1}_{-2.2})\times 10^{-3}$ at $z=0$ is an
estimate from the observed mean $B$-band emissivity of nearby galaxies
and the mean mass-to-light ratios in spheroids, disks, and irregular
galaxies (Fukugita, Hogan, \& Peebles 1998), while the data point
$\Omega_g=(5.0\pm1.2)\times10^{-4}$ at $z=0$ comes from observations
of the 21~cm emission of nearby galaxies (\S~2.2.6).  The results from
our models are plotted as the curves in each panel of Figure 14.  The
agreement with $\Omega_g(0)$ is by design, while the agreement with
$\Omega_s(0)$ is an indication that our adopted lower mass cutoff in
the IMF is reasonable.  Evidently, galaxies contain mostly stars at
$z\lesssim1$ and mostly gas at $z\gtrsim2$.  The bulk of the stars
seen today formed at relatively low redshifts: $34-48$\% since $z=1$,
$72-91$\% since $z=2$, $91-98$\% since $z=3$, and $97-99$\% since
$z=4$.

{\topinsert\null\vskip-12pt
  \psfig{file=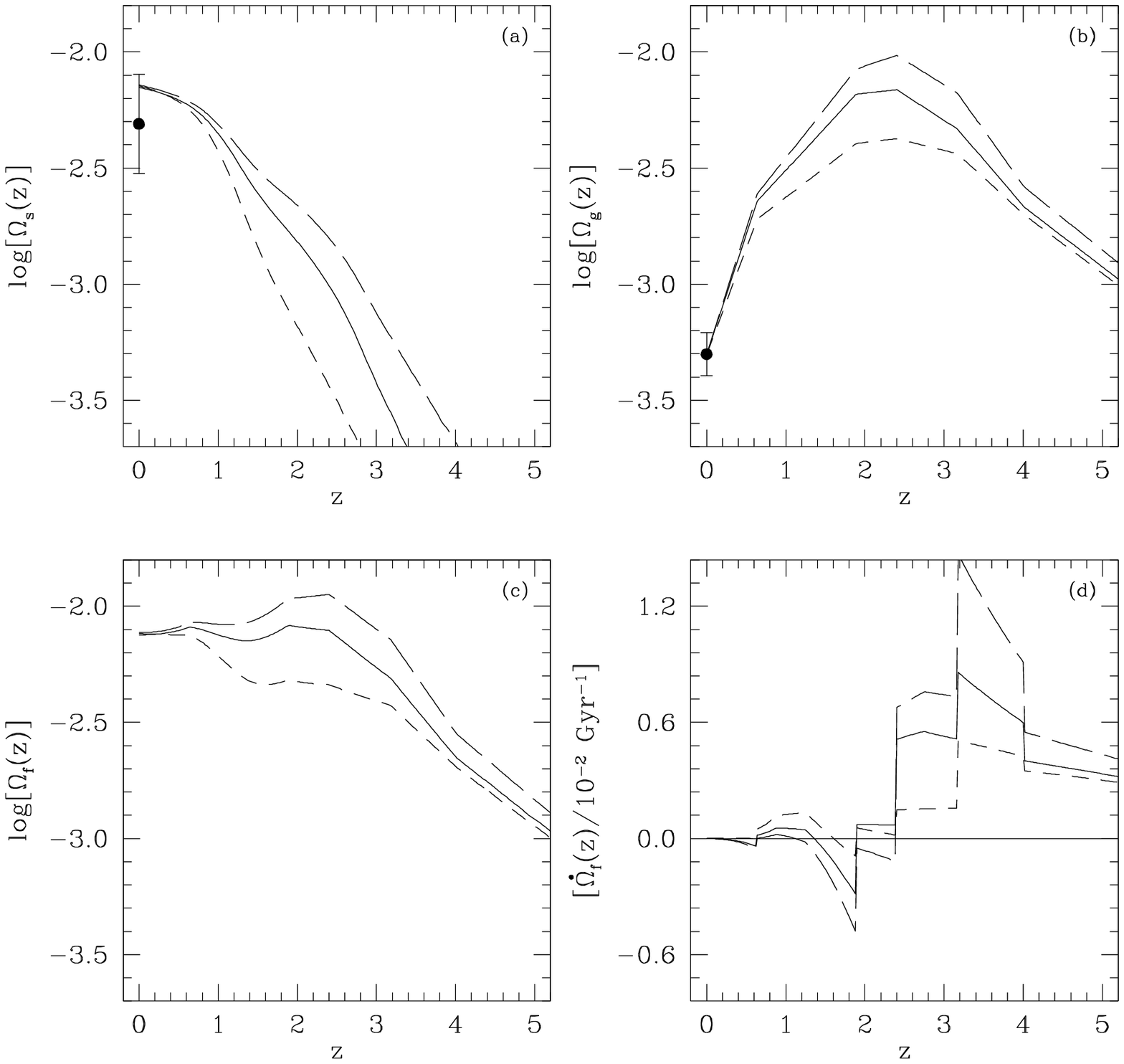,width=\h@size,silent=yes}
  \vskip12pt\edef\h@old{\the\hsize}\hsize=\h@size
  \vskip-48pt
  F{\sc IG}.~{14}.---{Evolution of (a) the comoving density of stars,
(b) the comoving density of interstellar gas, (c) the comoving density
of baryons in galaxies, and (d) the comoving rate of baryon flow
between the interstellar and intergalactic media.  The solid,
short-dash, and long-dash curves are the best, minimum, and maximum
solutions in our models.  The data point for $\Omega_s(0)$ is an
estimate from the mean blue luminosity density and mass-to-light ratio
of nearby galaxies (Fukugita \etal 1998), while the data point for
$\Omega_g(0)$ is inferred from 21~cm emission observations of nearby
galaxies (\S~2.2.6).}
  \vskip24pt\vskip-5.23433pt\vskip-0.7pt
  \hsize=\h@old\endinsert\vfuzz=0.5pt}

Although there are still significant uncertainties in these solutions,
they can be divided roughly into three periods: (1) At $z\gtrsim3$,
there is a rapid inflow from the intergalactic medium and hence a
build up of the interstellar gas within galaxies (Figs.~14b and 14c).
(2) At $1\lesssim z\lesssim3$, the interstellar gas content is high
(Fig.~14b), presumably triggering high rates of star formation
(Fig.~13).  This in turn causes the rapid accumulation of stars
(Fig.~14a) and heavy elements (Fig.~8), and strong emission from stars
and dust (Fig.~9).  (3) At $z\lesssim1$, the stellar content continues
to increase but at a slower rate, while the gas content (Fig.~14b),
star formation rate (Fig.~13), ultraviolet and optical stellar
emission (Fig.~9), and far-infrared dust emission (Fig.~9) all
decrease.  During this period, the interstellar gas is consumed by
star formation without much replenishment by inflow.  Our results
therefore suggest that, on average, galaxies spend $13-18$\% of their
lives (defined by the present age of the universe) in a ``growth''
period, $23-25$\% of their lives in a ``working'' period, and
$64-57$\% of their lives in a ``retirement'' period (for $\Lambda=0$
and $0.1\le q_0 \le0.5$).

It appears from our models that the most significant inflow occurred
at $z\approx3-4$ (Fig.~14d).  This inferrence, however, depends
crucially on the estimates of $\Omega_{{\rm HI}o}$ from damped
Ly$\alpha$ surveys at the highest redshifts (Fig.~7a).  If the
observed HI content were underestimated at $z\gtrsim3$, the peak in
the inflow could be pushed to an earlier epoch.  Since the
interstellar metallicity is inversely proportional to the comoving
density of gas, the observed metallicity at $z\approx3- 4$ also
provide a constraint on the peak in the inflow.  Therefore, better
determinations of the comoving density of HI and mean metallicity in
the damped Ly$\alpha$ systems at $z\gtrsim3$ would help reduce the
uncertainties in the epoch when the bulk of the interstellar gas was
assembled in galaxies.

In most models based on hierarchical clustering, including the cold
dark matter model, galaxies form relatively late.  The exact evolution
depends on a large number of processes, including the merging of dark
matter halos, the gravitational heating and radiative cooling of gas
within them, the formation of stars, the feedback of stellar mass,
energy, and metals to the interstellar medium, and so forth.
Semianalytical models treat these processes by simple,
physically-motivated recipes; and with suitable choices of parameters,
they are found to be consistent with many of the observed properties
of galaxies (White \& Frenk 1991; Kauffmann, White, Guiderdoni 1993;
Cole \etal 1994; Somerville \& Primack 1998). In some semianalytical
models, the comoving rate of star formation and the comoving density
of cold gas rise to peak values at $z\approx1-2$ and then decline at
lower redshifts (Kauffmann 1996; Baugh \etal 1998).  These models are
therefore qualitatively consistent with our models of cosmic chemical
evolution and the observations on which they are based.  The evolution
of $\dot\Omega_s$ and $\Omega_g$ in the semianalytical models is,
however, milder than the observed evolution.  It is not yet clear
whether or not this represents a significant discrepancy.

\subsection {4.5. Chemical Enrichment of the Intergalactic Medium}

The chemical enrichment of the intergalactic medium could be the
result of an outflow of metal-enriched gas from galaxies or
nucleosynthesis in objects much smaller and more uniformly distributed
than galaxies (e.g., pre-galactic stars or star clusters).  We explore
in this section the possibility of outflows from galaxies.  To do so,
we write the comoving density of metals in the intergalactic medium as
$$
\Omega_{m{\rm IGM}}(z)=\int_z^\infty dz'\biggl|{dt\over dz'}\biggr|
\dot\Omega_{\rm O}(z')Z(z'), \eqno(45)
$$
where $\dot\Omega_{\rm O}$ is the comoving rate of outflow from
galaxies and $Z$ is the mean metallicity of the outflowing gas.  The
net flow rate discussed in \S~2.1 is simply $\dot\Omega_f=\dot
\Omega_{\rm I}-\dot \Omega_{\rm O}$, where $\dot\Omega_{\rm I}$ is the
comoving rate of inflow.  Both $\dot\Omega_{\rm O}$ and $\dot
\Omega_{\rm I}$ must be positive, whereas $\dot\Omega_f$ can be
positive or negative.  We assume, as a good approximation, that
$\dot\Omega_{\rm O}=-\dot\Omega_f$ for $\dot\Omega_f<0$ (by ignoring
any inflow if the net flow is dominated by outflow) and
$\dot\Omega_{\rm O}=\eta \dot\Omega_f$ for $\dot\Omega_f>0$ with
$\eta\ll1$ (to allow for some outflow if the net flow is dominated by
inflow).  A small but non-zero value of $\eta$ only affects equation
(3) in \S~2.1.  We have checked that all of the results presented in
\S~3 are nearly the same for $\eta\lesssim0.3$, with a maximum
difference of less than 20\% in $Z$.  In this case, we can make rough
estimates of the chemical enrichment of the intergalactic medium from
our solutions for the net rate of baryon flow.

Figure~15 shows the mean comoving density of metals in the
intergalactic medium, resulting from the outflow of metal-enriched gas
from galaxies in our models.  The three curves represent our three
solutions with $\eta=0.2$.\break

\vskip19pt
\psfig{file=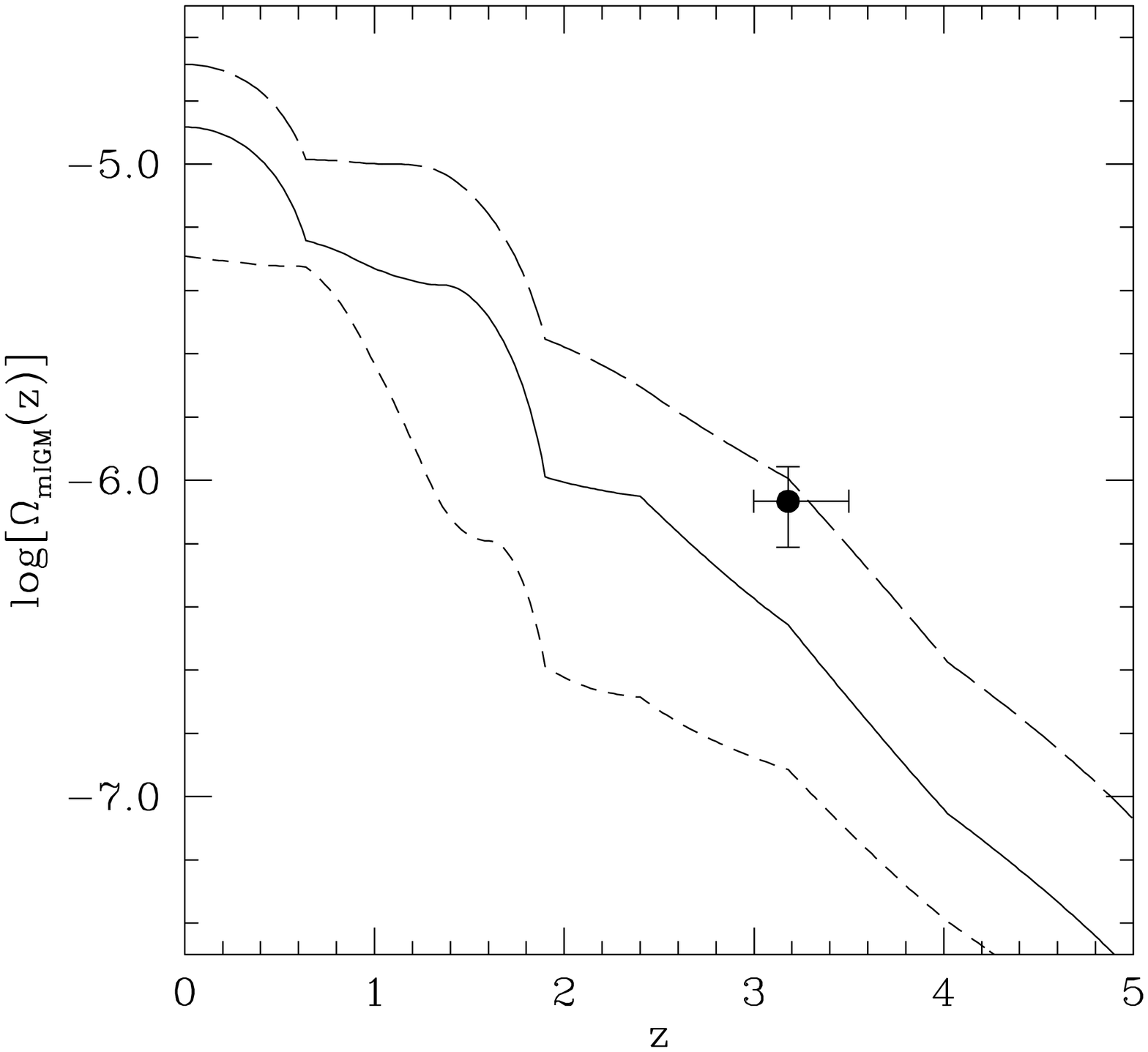,width=3.23in}
\vskip-12pt
F{\sc IG}.~15.---Mean comoving density of metals expelled from
galaxies into the intergalactic medium as a function of redshift.  The
solid, short-dash, and long-dash curves are the best, minimum, and
maximum solutions in our models.  The data point at $z=3.2$ was
derived by Songaila (1997) from observations of C~IV and Si~IC
absorption lines in the Ly$\alpha$ forest with some ionization
corrections and assumptions about relative abundances.

\noindent
The data point at $z=3.2$ was derived by
Songaila (1997) from observations of C~IV and Si~IV absorption lines
in the Ly$\alpha$ forest (corresponding roughly to $N_{\rm HI}\lesssim
10^{17}~{\rm cm}^{-2}$) with some modest ionization corrections and
assumptions about the relative abundances of different elements.
Although the data and models are both uncertain, the rough agreement
indicates that even a small outflow of metal-enriched gas from
galaxies could account for the observed metal content of the
Ly$\alpha$ forest.  A comparison of Figure~15 with Figure~8 suggests
that the comoving density of intergalactic metals is typically an
order of magnitude smaller than that of interstellar metals and that
the two are comparable only at $z\lesssim0.4$.  This is nearly
independent of the relative outflow rate for $\eta\lesssim0.3$.
Galaxies probably retained most of their metals up until now, but
there could be almost as many metals in the intergalactic medium as in
the interstellar media of galaxies at $z=0$.

The global mean metallicity of the intergalactic\break
\noindent
medium is $\langle
Z_{\rm IGM}\rangle=\Omega_{m{\rm IGM}}/\Omega_{g{\rm IGM}}$, where
$\Omega_{g{\rm IGM}}$ is the total gas content in the intergalactic
medium.  Adopting $\Omega_{g{\rm IGM}}\approx0.02$ from Big Bang
nucleosynthesis (Fukugita \etal 1998), we estimate $\log \langle
Z_{\rm IGM}/Z_\odot\rangle\approx-1.5$, $-2.6$, and $-3.6$ at $z=0$,
2, and 4, respectively.  We emphasize, however, that there could be a
large dispersion around the mean, depending on whether and how much
the metals expelled from galaxies are mixed with the intergalactic
medium.  If the metals were mixed completely with all of the
intergalactic gas, the metallicity would be the same everywhere, i.e.,
$Z_{\rm IGM}=\langle Z_{\rm IGM}\rangle$.  If the metals (and gas)
were not mixed at all in the intergalactic medium, the metallicity
would fluctuate between $Z_{\rm IGM}=Z$ in regions where the ejected
gas is present and $Z_{\rm IGM}=0$ in regions where it is not present.
Some degree of inhomogeneity would seem to be the most likely
situation.

\section {5. CONCLUSIONS}

We have studied the global evolution of the stellar, gaseous, and
metal contents of galaxies, as well as the radiation they emit and the
baryon flow between the interstellar and intergalactic media.  Our
results are based on models of cosmic chemical evolution, with
observational inputs from damped Ly$\alpha$ absorption-line surveys
and optical imaging and redshift surveys, and measurements of the
cosmic infrared background.  The novel feature of our approach is a
self-consistent treatment of the obscuration, absorption, and emission
by dust.  We emphasize that all of our results pertain to the mean
properties of galaxies, averaged over comoving volumes large enough to
be fair samples of galaxies of all masses, sizes, and morphological
types.  Such a global approach has the advantage of simplicity, but
the disadvantage that many interesting questions concerning the
diversity of galaxies cannot be addressed.  In particular, our results
are completely independent of the unknown morphologies, sizes, and
space densities of the damped Ly$\alpha$ systems.

We find that the extragalactic far-infrared background detected by the
{\it COBE} DIRBE and FIRAS can be accounted for if much of the
radiation emitted by stars is absorbed and reemitted by small-scale
clumps of dust, possibly associated with individual stars.  The ratio
of the bolometric dust emission to the bolometric stellar emission in
our models increases from 36\% at $z=0$ to about 60\% at $z\approx 1$
and becomes uncertain at higher redshifts ($10-70$\% at $z\gtrsim 3$).
Of the total energy radiated by stars (integrated over all redshifts),
$41-46$\% is in the ultraviolet/optical/near-infrared part of
extragalactic background, while the remaining $59-54$\% is in the
mid-infrared/far-infrared/submillimeter part.  The absorption by dust
corrects the global rates of star formation inferred from optical
surveys upward by factors of 2.5, $3-4$, and $2-8$ at $z=0$, 1, and 3,
respectively.  From our models, it appears that galaxies may have
experienced a sustained period of intense star formation at $1\lesssim
z\lesssim3$ and somewhat less rapid star formation at $z\lesssim 1$
and $z\gtrsim3$, roughly correlated with the evolution of their
neutral gas content, although the evolution at $z\gtrsim3$ is quite
uncertain.  We note, however, that if the global rate of star
formation were to remain high beyond $z\approx3$, it would produce
more metals in the interstellar medium than are observed in the damped
Ly$\alpha$ systems, unless a significant outflow had occurred before
$z\approx3$.

The solutions presented here for the mean properties of galaxies are
based on relatively few input observations, but they match remarkably
well a wide variety of other observations, mostly within the quoted
$1\sigma$ errors of the data.  In particular, our models reproduce:
(1) the extragalactic background derived from galaxy counts at
$\lambda=0.2-2.2~\mu$m; (2) the rest-frame 0.44, 1.0, and 2.2~$\mu$m
emissivities from the available surveys of galaxies up to $z
\approx2$; (3) the infrared dust emissivities from nearby galaxies at
$\lambda=12$, 25, 60, and $100~\mu$m; (4) the observed mean abundance
of heavy elements in the damped Ly$\alpha$ systems at $0.4 \lesssim
z\lesssim3.5$; and (5) the global rates of star formation inferred
from H$\alpha$ and ISO mid-infrared surveys at $z\lesssim 1$ and
recent SCUBA submillimeter observations.  The metal enrichment history
of the intergalactic medium implied by our models is also consistent
with the observed mean comoving density of metals in the Lyman-alpha
forest at high redshifts.  Our solutions are robust at $z\lesssim2$
but uncertain at $z\gtrsim3$.  Major sources of uncertainties are the
optical opacity of grains and the small-scale clumpiness of dust in
the interstellar medium.  Both of these are likely to remain uncertain
for some time and hence will affect the interpretations of future
optical and infrared observations as well.

Our results indicate that many different observations of galaxies, by
means of quasar absorption lines, optical imaging and redshift
surveys, and extragalactic background measurements, can be brought
together to provide a coherent picture of the global evolution of
galaxies over much of the Hubble time.  The models presented here
suggest that the process of galaxy formation may have undergone
different evolutionary phases: (1) an early period of significant
inflow to assemble interstellar gas at $z\gtrsim3$; (2) a subsequent
period of intense star formation and chemical enrichment at $1\lesssim
z\lesssim 3$; and (3) a recent period of decline in the gas content,
star formation rate, and the ultraviolet and optical stellar
emissivity, and far-infrared dust emissivity at $z\lesssim1$.
Currently, there are still large uncertainties in this picture,
especially at $z\gtrsim3$.  Future surveys that aim to trace the
rest-frame near-ultraviolet emissivity of stars, the rest-frame
far-infrared emissivity of dust, and the global properties of
interstellar gas, heavy elements, and dust beyond $z\approx3$ could
reduce these uncertainties and thus advance our knowledge of the
global evolution of galaxies.

\vskip24pt

We thank Rick Arendt, St\'ephane Charlot, Mark Dickinson, Eli Dwek,
Piero Madau, and Nino Panagia for helpful discussions.  We are grateful
to St\'ephane Charlot for providing his latest stellar population
synthesis models, to Lisa Storrie-Lombardi for communicating the
redshift path information of the damped Ly$\alpha$ sample, and to Dale
Fixsen for making the unpublished errors in the FIRAS data available to
us.  M.G.H and Y.C.P acknowledge {\it COBE} project support from NASA
grant NAG 5-3899. 

\vskip12pt

\section {REFERENCES}
\reference Armand, C., Milliard, B., \& Deharveng, J.M. 1994, A\&A,
             284, 12
\reference Barger, A.J., Cowie, L.L., Sanders, D.B., Fulton, E.,
             Taniguchi, Y., Sato, Y., Kawara, K., \& Okuda, H. 1998,
             Nature, 394, 248
\reference Baugh, C.M., Cole, S., Frenk, C.S., \& Lacey, C.G. 1998,
             ApJ, 498, 504
\reference Biller, S.D., Buckley, J., Burdett, A., Gordo, J.B.,
             Carter-Lewis, D.A., Fegan, D.J., Findley, J., Gaidos, J.A.,
             Hillas, A.M., Krennrich, F., Lamb, R.C., Lessard, R.,
             McEnery, J.E., Mohanty, G., Quinn, J., Rodgers, A.J.,
             Rose, H.J., Samuelson, F., Sembroski, G., Skelton, P.,
             Weekes, T.C., \& Zweerink, J. 1998, Phys. Rev. Lett.,
             80, 2992
\reference Blain, A.W., Smail, I., Ivison, R.J., \& Kneib, J.-P. 1999,
             MNRAS, 302, 632
\reference Boiss\'e, P., Le Brun, V., Bergeron, J., \& Deharveng, J.-M.
             1998, A\&A, 333, 841
\reference Bowyer, S. 1991, ARA\&A, 29, 59
\reference Bruzual, A., \& Charlot, S. 1998, in preparation
\reference Calzetti, D., \& Heckman, T.M. 1999, ApJ, in press,
             astro-ph/9811099
\reference Cole, S., Aragon-Salamanca, A., Frenk, C.S., Navarro, J.F.,
             \& Zepf, S.A. 1994, MNRAS, 271, 781
\reference Connolly, A.J., Szalay, A.S., Dickinson, M., Subbarao,
             M.U., \& Brunner, R.J. 1997, ApJ, 486, L11
\reference Cowie, L.L., Songaila, A., Hu, E.M., \& Cohen, J.G. 1996,
             ApJ, 112, 839
\reference D\'esert, F.X., Boulanger, F., Puget, J.L. 1990, A\&A, 237,
             215
\reference Draine, B.T., \& Lee, H.M.  1984, ApJ, 285, 89
\reference Dube, R.R., Wickes, W.W., \& Wilkinson, D.T.  1979, ApJ,
             232, 333
\reference Dwek, E. 1998, ApJ, 501, 643
\reference Dwek, E., \& Arendt, R.G. 1998, ApJ, 508, L9
\reference Dwek, E., Arendt, R.G., Hauser, M.G., Fixsen, D., Kelsall,
             T., Leisawitz, D., Pei, Y.C., Wright, E.L., Mather, J.C.,
             Moseley, S.H., Odegard, N., Shafer, R., Silverberg, R.F.,
             \& Weiland, J.L.  1998, ApJ, 508, 106
\reference Ellis, R.S., Colless, M., Broadhurst, T., Heyl, J., \&
             Glazebrook, K.  1996, MNRAS, 280, 235
\reference Fall, S.M. 1998, in Hubble Deep Field, ed. M. Livio, S.M.
             Fall, \& P. Madau (Cambridge: Cambridge University Press),
             163
\reference Fall, S.M., Charlot, S., \& Pei, Y.C.  1996, ApJ, 464, L43
\reference Fall, S.M., \& Pei, Y.C.  1993, ApJ, 402, 479
\reference Fixsen, D.J., Cheng, E.S., Gales, J.M., Mather, J.C., Shafer,
             R.A., \& Wright, E.L. 1996, ApJ, 473, 576
\reference Fixsen, D.J., Dwek, E., Mather, J.C., Bennett, C.L., \&
             Shafer, R.  1998, ApJ, 508, 123
\reference Flores, H., F. Hammer, F., Thuan, T., C\'esarsky, C., D\'esert,
             F.X., Omont, A., Lilly, S.J., Eales, S., Crampton, D., \&
             Le F\`evre, O. 1999, ApJ, in press, astro-ph/9811202
\reference Fukugita, M., Hogan, C.J., \& Peebles, P.J.E. 1998, ApJ,
             503, 518
\reference Gallego, J., Zamorano, J., Ar\'agon-Salamanca, A., \&
             Regg, M. 1995, ApJ, 455, L1
\reference Gardner, J.P., Sharples, R.M., Frenk, C.S., \& Carrasco,
             B.E. 1997, ApJ, 480, 99
\reference Ge, J., \& Bechtold, J. 1997, ApJ, 477, L73
\reference Glazebrook, K., Blake, C., Economou, F., Lilly, S., \&
             Colless, M. 1999, MNRAS, in press, astro-ph/9808276
\reference Hacking, P.B., \& Soifer, B.T. 1991, ApJ, 367, L49
\reference Hartwick, F.D.A., \& Schade, D.  1990, ARA\&A, 28, 437
\reference Hauser, M.G., Arendt, R.G., Kelsall, T., Dwek, E., Odegard,
             N., Weiland, J.L., Freudenreich, H.T., Reach, W.T.,
             Silverberg, R.F., Moseley, S.H., Pei, Y.C., Lubin, P.,
             Mather, J.C., Shafer, R.A., Smoot, G.F., Weiss, R.,
             Wilkinson, D.T., \& Wright, E.L.  1998, ApJ, 508, 25
\reference Hogan, C.J., Oliver, K.A., \& Scully, S.T. 1997, ApJ, 489,
             L119
\reference Holberg, J. 1986, ApJ, 311, 969
\reference Hughes, D., Serjeant, H., Dunlop, J., Rowan-Robinson, M.,
             Blair, A., Mann, R.G., Ivison, R., Peacock, J., Efstathiou,
             A., Gear, W., Oliver, S., Lawrence, A., Longair, M., 
             Goldschmidt, P., Jenness, T. 1998, Nature, 394, 241
\reference Kashlinsky, A., Mather, J.C., Odenwald, S. 1996, ApJ, 473, L9
\reference Kauffmann, G. 1996, MNRAS, 281, 475
\reference Kauffmann, G., White, S.D.M., \& Guiderdoni, B. MNRAS, 264,
             201
\reference Kulkarni, V.P., S.M., Fall, \& Truran, J.W. 1997, ApJ, 484,
             L7
\reference Lanzetta, K.M., Wolfe, A.M., \& Turnshek, D.A.  1995, ApJ,
             440, 435
\reference Lanzetta, K.M., Wolfe, A.M., Turnshek, D.A., Lu, L., McMahon,
             R.G., \& Hazard, C.  1991, ApJS, 77, 1
\reference Le Brun, V., Bergeron, J., Boiss\'e, P., \& Deharveng, J.M.
             1997, A\&A, 321, 733
\reference Leitherer, C., Furguson, H.C., Heckman, T.M., \& Lowenthal,
             J.D.  1995, ApJ, 454, L19
\reference Levshakov, S.A, Chaffee, F.H., Foltz, C.B., \& Black, J.H. 
             1992, A\&A, 262, 385
\reference Lilly, S.J., Eales, S.A., Gear, W.K.P., Hammer, F., Le Fevre,
             O., Crampton, D., Bond, J.R., \& Dunne, L. 1999, ApJ, in press,
             astro-ph/9901047
\reference Lilly, S.J., Le F\'evre, O., Hammer, F., \& Crampton, D. 
             1996, ApJ, 460, L1
\reference Lilly, S.J., Tresse, L., Hammer, F., Crampton, D., \& Le
             F\'evre, O.  1995, ApJ, 455, 108
\reference Luck, R.E., \& Lambert, D.L.  1992, ApJS, 79, 303
\reference Madau, P., Ferguson, H.C., Dickinson, M.E., Giavalisco, M.,
             Steidel, C.C., \& Fruchter, A.  1996, MNRAS, 283, 1388
\reference Madau, P., Pozzetti, L., \& Dickinson, M. 1998, ApJ, 498, 106
\reference Natta, A., \& Panagia, N. 1984, ApJ, 287, 228
\reference Ostriker, J.P., \& Heisler, J. 1984 ApJ, 278, 1
\reference Pei, Y.C.  1992, ApJ, 395, 130
\reference Pei, Y.C.  1995, ApJ, 438, 623
\reference Pei, Y.C., \& Fall, S.M.  1995, ApJ, 454, 69
\reference Pei, Y.C., Fall, S.M., \& Bechtold, J.  1991, ApJ, 378, 6
\reference Perna, R., Loeb, A., \& Bartelmann, M. 1997, ApJ, 488, 550
\reference Pettini, M., King, D.L., Smith, L.J., \& Hunstead, R.W. 
             1997a, ApJ, 478, 536
\reference Pettini, M., Smith, L.J., Hunstead, R.W., \& King, D.L. 
             1994, ApJ, 426, 79
\reference Pettini, M., Smith, L.J., King, D.L., \& Hunstead, R.W. 
             1997b, ApJ, 486, 665
\reference Pozzetti, L., Madau, P., Ferguson, H.C., Zamorani, G., \&
             Bruzual, G.A. 1998, MNRAS, 298, 1133
\reference Puget, J.-L., Abergel, A., Bernard, J.-P., Boulanger, F.,
             Burton, W.B., D\'esert, F.-X., \& Hartmann, D. 1996, A\&A,
             308, L5
\reference Rao, S., \& Briggs, F.  1993, ApJ, 419, 515
\reference Reiss, A.G., \etal 1998, AJ, 116, 1009
\reference Schelgel, D.J., Finkbeiner, D.P., \& Davis, M. 1998, ApJ,
             500, 525
\reference Shaver, P.A., Wall, J.V., Kellermann, K.I., Jackson, C.A.,
             \& Hawkins, M.R.S. 1996, Nature, 384, 439
\reference Smette, A., Claeskens, J.-F., \& Surdej, J. 1997, New
             Astronomy, 2, 53
\reference Soifer, B.T., \& Neugebauer, G. 1991, AJ, 101, 354
\reference Somerville, R.S., \& Primack, J.R. 1998, MNRAS, submitted,
             astro-ph/9802268
\reference Songaila, A. 1997, ApJ, 490, L1
\reference Stanev, T., \& Franceschini, A. 1998, ApJ, 494, 159
\reference Steidel, C.C., Adelberger, K.L., Giavalisco, M., Dickinson,
             M., \& Pettini, M. 1999, ApJ, in press, astro-ph/9811399
\reference Steidel, C.C., Giavalisco, M., Pettini, M., Dickinson, M.,
             \& Adelberger, K.L. 1996, ApJ, 462, L17
\reference Storrie-Lombardi, L.J., Irwin, M.J., \& McMahon, R.G. 
             1996a, MNRAS, 282, 1330
\reference Storrie-Lombardi, L.J., McMahon, R.G., \& Irwin, M.J. 
             1996b, MNRAS, 283, L79
\reference Tinsley, B.M. 1980, Fund. Cosmic Phys., 5, 287
\reference Toller, G.N.  1983, ApJ, 266, L79
\reference Tresse, L., \& Maddox, S.J. 1998, ApJ, 495, 691
\reference Vladilo, G. 1998, ApJ, 493, 583
\reference White, S.D.M., \& Frenk, C.S. 1991, ApJ, 379, 52
\reference Williams, R.E., Blacker, B., Dickinson, M., van Dyke Dixon,
             W., Ferguson, H.C., Fruchter, A.S., Giavalisco, M., 
             Gilliland, R.L., Heyer, I., Katsanis, R., Levey, Z., Lucas,
             R.A., McElroy, D.B., Petro, L., \& Postman, M.  1996, AJ, 
             112, 1335
\reference Wolfe, A.M., Turnshek, D.A., Smith, H.E., \& Cohen, R.D. 
             1986, ApJS, 61, 249
\reference Wolfe, A.M., Lanzetta, K.M., Foltz, C.B., \& Chaffee, F.H. 
             1995, ApJ, 454, 698
\reference Young, J.S., \& Scoville, N.Z.  1991, ARA\&A, 29, 581
\reference Zwaan, M.A., Briggs, F.H., Sprayberry, D., \& Sorar, E. 1997,
             ApJ, 490, 173
\end\bye